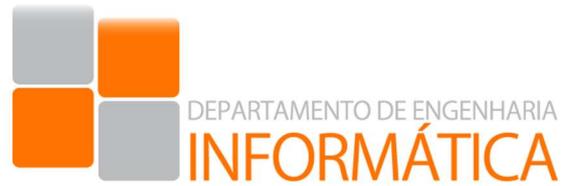

# A Scalable Clustered Architecture for Cyber-Physical Systems

ISEP – Instituto Superior de Engenharia do Porto

## 2022 / 2023

**1200852 Bernardo Magalhães Amaral Cabral**

**ISEP** INSTITUTO SUPERIOR DE ENGENHARIA DO PORTO

# A Scalable Clustered Architecture for Cyber-Physical Systems

ISEP – Instituto Superior de Engenharia do Porto

## 2022 / 2023

**1200852 Bernardo Magalhães Amaral Cabral**

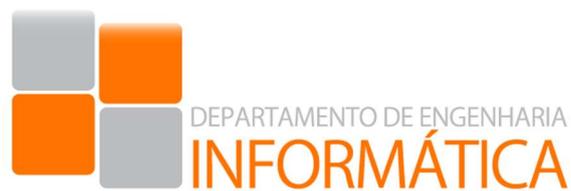

# Degree in Informatics Engineering

## June 2023

ISEP Advisor: Luis Miguel Moreira Lino Ferreira

*"In the end, we only regret the chances we didn't take."*

**Lewis Carroll**

# Acknowledgments

Firstly, I would like to express my gratitude to my family, especially my parents and brother, as well as my grandparents, aunt, and godparents, for the support and sacrifices they have made for me to reach where I am today and for shaping me into the person I am. I would like to express my appreciation to my girlfriend, who has been by my side since the end of high school, always supporting me and helping me overcome various challenges to achieve my goals. In addition to that, I want to express my gratitude to my tutor Professor Célia Correia during the high school, who has always encouraged me to pursue the field of computer science and follow my goals, and has been there every step of the way, supporting and following my achievements.

I would also like to express my gratitude to my internship advisor, Professor Luis Miguel Moreira Lino Ferreira, for always being available to help and guide me throughout the project. It was thanks to him that I had the opportunity to carry out this project.

Furthermore, I want to express my gratitude to ISEP for providing me with the tools to develop my technical skills and achieve my goals.

And last, but certainly not least, I want to thank two of my research colleagues from ISEP, Tiago Carlos Caló Fonseca and Pedro Silva Costa, for working with me and helping me in whatever was necessary, and for all the moments we shared throughout the project.

# Abstract

Cyber-Physical Systems (CPS) play a vital role in the operation of intelligent interconnected systems. CPS integrates physical and software components capable of sensing, monitoring, and controlling physical assets and processes.

However, developing distributed and scalable CPSs that efficiently handle large volumes of data while ensuring high performance and reliability remains a challenging task. Moreover, existing commercial solutions are often costly and not suitable for certain applications, limiting developers and researchers in experimenting and deploying CPSs on a larger scale.

The development of this project aims to contribute to the design and implementation of a solution to the CPS challenges. To achieve this goal, the Edge4CPS system was developed.

Edge4CPS system is an open source, distributed, multi-architecture solution that leverages Kubernetes for managing distributed edge computing clusters. It facilitates the deployment of applications across multiple computing nodes. It also offers services such as data pipeline, which includes data processing, classification, and visualization, as well as a middleware for messaging protocol translation.

# Index

# Table of Figures

# Table of Tables

# Table of Code Snippets

# Notation and Glossary

| | |
|---|---|
| **ISEP** | Instituto Superior de Engenharia do Porto |
| **Edge4CPS** | Proposed system in the thesis, which was developed throughout the internship. |
| **API** | Application Programming Interface. A set of clearly defined methods of communication between various software components |
| **UI** | Throughout the report, the term UI "user interface" will also be used to refer to the "web interface. |
| **CPS** | Cyber Physical System |
| **UML** | Unified Modelling Language, a software modelling language |
| **ARM** | Advanced RISC Machine |
| **IoT** | Internet of Things |
| **AI** | Artificial Intelligence |
| **CPU** | Central Processing Unit |
| **GPU** | Graphics Processing Unit |
| **JSON** | Java Script Object Notation |
| **HTTP** | Hypertext Transfer Protocol |
| **SSH** | Secure Shell |

# 1 Introduction

This initial chapter provides a broad perspective on the project, including its context, objectives, approach, and thesis organization.

## 1.1 Project Context

Cyber-Physical Systems (CPS) consolidate the physical and software components capable of sensing, monitoring, and controlling physical assets or processes. These systems have the tendency of being distributed and comprised by a large set of different subsystems (Gunes et al., 2014). Each running on specific architectures, which span from IoT devices (e.g., based on 8-bit microcontrollers) to high end cloud systems (usually based on 64-bit architectures).

However, developing distributed and scalable CPSs that can efficiently handle large amounts of data, while delivering high performance and ensuring high levels of reliability, remains a hard task. Furthermore, as we will explore in Section II, the commercial solutions available in the market are not interesting for some applications due to their high costs. This makes it difficult for developers and researchers to experiment with and deploy CPSs on a larger scale. To overcome this challenge, it is imperative to extend support for multiple architectures in particularly at edge or fog level in an open-source solution available to everyone. At the same time, these systems must also provide developers with an easy-to-use environment where they can rapidly deploy their applications.

Given the current scenario of the CPS and IoT industries, architectures like ARM and RISC-V have gained popularity due to their ability to offer high-performance at a low-cost (Pa et al., 2015). As such, an edge system capable of supporting distributed applications on multiple architectures can better use existing resources and easily support the distribution of computing tasks, taking advantage of the unique strengths and limitations of each hardware type. By adopting such an approach, it is possible, for example, to optimize the overall performance and reduce latency, facilitating more effective processing of large volumes of data. Furthermore, with the rise of Artificial Intelligence (AI) applications (Deng, 2018), it is crucial that the system recognizes architectures that are specifically designed for AI, such as those used in dedicated graphics cards, as these are more efficient and faster on supporting those kinds of applications.

Considering this context, we present the Edge4CPS, an open source distributed multi-architecture solution supported by Kubernetes (Burns, 2018) for distributed edge computing clusters. Edge4CPS facilitates a rapid and scalable deployment of applications on multiple computing nodes. Additionally, Edge4CPS includes a variety of available services for the user to utilize, such as multiple messaging protocol brokers, a multi-protocol translator, and a data pipeline for data processing, handling, classification, and visualization.

### 1.1.1 Internship Goals

The goals of this project can be categorized into three components: i) the first component involves implementing the cluster and its critical components; ii) the second component focuses on developing deployment and management modules for the cluster; and iii) the final component entails implementing a parallel distributed pipeline for data processing and handling within the cluster. The goals will be designated by (G).

*G 1.* Implement the system using Kubernetes with a focus on scalability, maintainability, and performance.

*G 2.* Ensure the system can support multiple applications and services with diverse requirements in a clustered environment.

*G 3.* Support the design, implementation, and test of a middleware, within the cluster's containerized environment, that is capable of translating messages between multiple messaging protocols, like Kafka and MQTT.

*G 4.* Design, develop, and test an API that can handle requests to the Kubernetes Framework within the master node, making it easier to deploy applications and services.

*G 5.* Develop a user-friendly web interface that implements the existing API, simplifying its complexity and offering users a graphical interface to manage the deployment of applications and services.

*G 6.* Establish a data pipeline within the cluster that enables efficient processing, handling, classification, and visualization of data.

### 1.1.2 Approach

It can be said that the approach was influenced by both the duration of the internship and the specific project in which it was implemented, in this case, the Ferrovia 4.0 project ("Ferrovia40 - The Official Website of Ferrovia 40.", 2023).

The adopted approach begins by identifying the requirements for the Edge4CPS system. After this initial process, the next step involves researching similar solutions and technologies that can be utilized and determining which ones are most suitable for the respective system. No specific technology was specified in the requirements, so after conducting an analysis, it was decided to use JavaScript with the React library for the frontend. Python was chosen for the backend implementation, while Kubernetes and Docker were selected for creating a cluster of nodes and setting up the containerized environment.

### 1.1.3   Contributions from this project

The Edge4CPS system is an open source (Bernardo Cabral GitHub Repository,2023), distributed, multi-architecture solution that facilitates the seamless deployment of applications within the system. Additionally, it provides pre-configured services that are readily available to users.

Since it is an open-source solution, Edge4CPS provides an alternative to the use of proprietary large cloud platforms for users who wish to implement the system.

The presence of an API and a UI to abstract the Kubernetes framework and simplify the deployment of applications can bring significant advantages to less experienced users who need to utilize the cluster, as in the case of users from other work packages on the Ferrovia 4.0 project. In addition to that, this system is designed to be used with low-cost devices, providing the opportunity for devices with limited processing capabilities to be utilized within the system.

The Edge4CPS system is designed to contribute to projects of varying sizes, supporting a wide range of deployment requirements. For small-scale projects, Edge4CPS offers a cost-effective solution that can be easily deployed on a limited number of edge nodes. It enables distributed computing capabilities, reduces latency, and optimizes resource utilization, all while providing a straightforward deployment process.

On the other hand, for large-scale projects, Edge4CPS provides scalability and flexibility to accommodate extensive deployments with numerous edge nodes. The system's ability to distribute workloads across multiple nodes and parallelize data processing services, which can handle multiple messaging protocols, ensures efficient resource utilization and improved performance at scale. This dynamic capability makes Edge4CPS suitable for complex and demanding applications that require significant processing power, real-time data analytics, and reliable operations in large-scale environments.

## 1.2 Project Planning

In this subsection, we will present the development process from the beginning to the completion of report writing, including the duration of each stage. Figure 1 demonstrates the Gantt chart. Initially, the research and exploration of existing technologies and solutions were conducted, along with the documentation of the state-of-the-art and the analysis of system requirements and design. Following a thorough study and preliminary analysis, the solution was implemented, encompassing all the essential components for the system's functionality. Throughout the implementation phase, testing was conducted concurrently with the implementation to ensure the proper functioning of the system. Throughout the project, an Agile software development approach was applied. As can be observed, the Agile development model (Edeki, 2015) was followed, starting with a research period and analysis and design of the system. This was followed by concurrent testing and implementation.

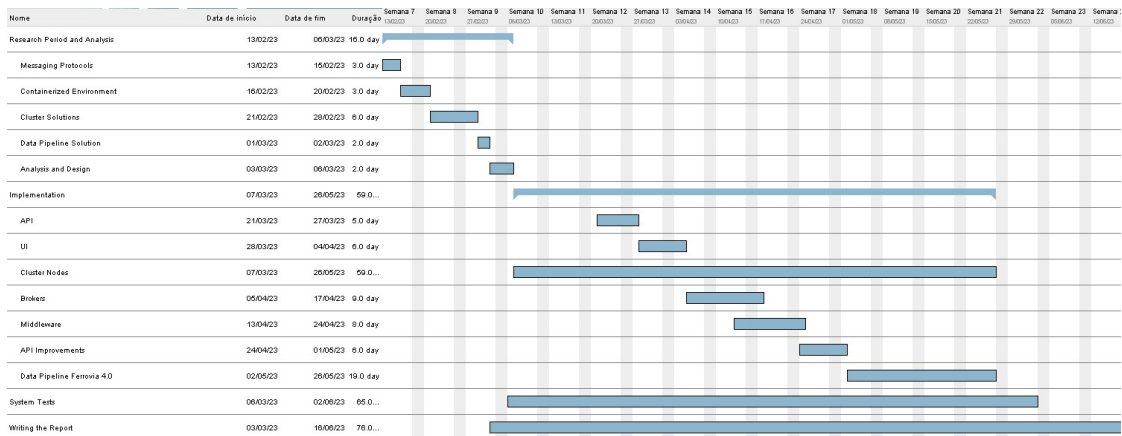

Figure 1 Gantt Diagram

### *1.2.1.1 Meetings*

The project meetings with the supervisor primarily took place in person at the meeting office of ISEP. For the few online meetings, the Teams tool was utilized. Similarly, the consortium meetings were also conducted online using the Teams platform. The meetings focused on discussing the project's progress, addressing challenges that arose, reviewing its requirements, and exploring solutions put forth by other partners.

## 1.3 Report Structure

This section provides an overview of the report's organization and a brief summary of the contents of each chapter and subchapter.

- **Chapter 2 - State of the Art:** This chapter aims to present the origin and theoretical framework of the project, as well as existing technologies. Additionally, after introducing similar projects, it will provide a brief comparison highlighting the differences with Edge4CPS.

- **Chapter 3 - Technical Description:** In this chapter, the requirements and preliminary analysis of the project will be discussed. This includes both functional and non-functional requirements, as well as the presentation of user stories and involved actors. Finally, a domain analysis will be conducted.

- **Chapter 4 - Architectural Design:** In the design chapter, a detailed overview of the decisions that have been made will be provided. It will also explore the various components of Edge4CPS and document its architecture.

- **Chapter 5 - Implementation:** In the implementation chapter, it provides an overview of the description of each component of the Edge4CPS system and how they interact with the system.

- **Chapter 6 -** In this chapter, all the tests conducted on each component of the system will be presented and discussed.

- **Chapter 7 - Conclusion:** This chapter will provide an overview of the implemented system, including a discussion on potential future improvements. It will then present the objectives that were successfully achieved as planned at the beginning of the project. Finally, it will provide insights into the internship experience and the proposed project.

# 2  State of the Art

The following chapter will allow the reader to understand the origin and theoretical standpoint of the project and existing technologies. Therefore, it will be necessary to begin by discussing the project context before delving into existing technologies and related projects.

## 2.1  Project Context

In today's world, centralized computing systems no longer meet the demands of modern applications and services. As a result, scalable, distributed systems capable of supporting efficient data processing, while simplifying the deployment of services, have become essential (Gunes et al., 2014). Cloud, fog, and edge computing have emerged as potential solutions to address these challenges (Kimovski et al., 2021).

- **Cloud computing:** When it comes to the cloud, computing is performed remotely from devices like sensors and other end devices. This type of computation is highly scalable, but it may introduce latency and is typically reserved for non-real time use cases. Additionally, it demands substantial computational power to operate effectively.

- **Fog computing:** In contrast, fog computing is located between the cloud and the edge. It works directly with the cloud, handling data that does not need to be processed on the go. However, fog computing is closer to the edge. If necessary, fog computing also has the ability to process and analyse data, providing a rapid response with latency but lower than that of the cloud.

- **Edge computing** Edge computing is the type of computing that interacts most closely with end devices. It can perform real-time processing as it has the lowest latency of the three types of computing. It also allows for computing and storage of a limited volume of data directly on devices, applications, and edge gateways.

Table 1 Types of Systems

|  | Cloud | Fog | Edge |
|---|---|---|---|
| **Latency** | *Highest* | *Medium* | *Lowest* |
| **Scalability** | *Easy to scale* | *Within network* | *Hard to scale* |
| **Distance** | *Far from edge* | *Close to Edge* | *At the Edge* |
| **Computing Power** | *High* | *Limited* | *Limited* |

Table 1 presents a comprehensive overview of different metrics and provides a comparative analysis of performance scores for cloud, fog, and edge computing ("Cloud, Fog, Edge Computing & IoT: Exploring the Intersection." Digiteum, 2023) across these metrics.

Although, our solution can also be deployed to support fog and cloud computing, in this project, we will focus on edge computing, once the Ferrovia 4.0 use case requires data processing in real time with low latency.

## 2.2  Clustering Solutions

As mentioned above, the system will have to be scalable, so it will be implemented in a cluster to support multiple nodes with different roles and have the ability to expand when necessary. When it comes to implementing clustering for systems, there are several technologies available to choose from. Some of the popular options include Kubernetes, and Docker Swarm (Marathe et al., 2019).

### 2.2.1  Docker Swarm

Docker Swarm ("Docker Swarm: How Swarm Mode Works - Nodes.", 2023) is an open-source container orchestration platform built and maintained by Docker. The installation process is very straightforward and user friendly, in contrast to other cluster solutions. With Docker (Anderson, 2015) already in the system there is no required configuration changes for Docker Swarm. One major disadvantage of Docker Swarm is that its automation capabilities are not as robust as Kubernetes-based container orchestration platforms (Burns & Tracey, 2018). Docker Swarm is typically preferred for less complex projects.

Docker Swarm consists of three main components, Swarm Manager, Swarm Nodes, and Overlay Networks.

The Swarm Manager is responsible for controlling the entire swarm and orchestrating the deployment of services across the cluster. It manages the cluster state, schedules tasks, and coordinates the communication between nodes. The Swarm Manager also exposes an API that can be used to interact with the cluster itself.

Swarm Nodes are the worker machines in the cluster. They run Docker Engine and execute the tasks and services assigned to them by the Swarm Manager. Nodes can be either manager nodes or worker nodes, depending on their role. Manager nodes participate in the control plane and handle cluster management tasks, while worker nodes are responsible for running containerized applications.

The last component is the Overlay networks, which provide a way for containers running on different nodes to communicate with each other. Swarm uses a built-in overlay network driver to create a distributed network across the swarm. This enables containers to seamlessly communicate with each other, regardless of the node they are running on. Fig. 2 presents a view of Docker Swarm components.

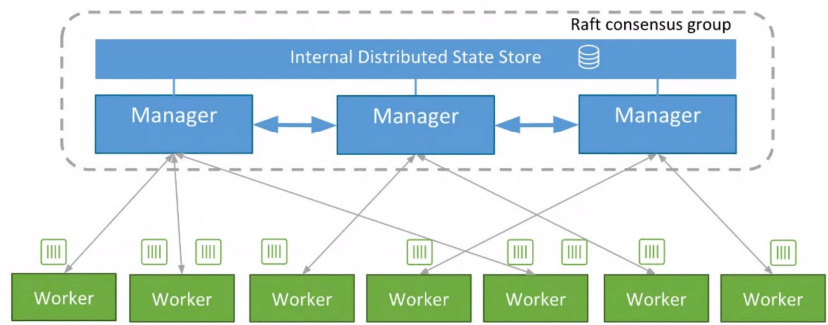

Figure 2 Docker Swarm architecture components

### 2.2.2 Kubernetes

Kubernetes (Burns & Tracey, 2018) is an open-source container orchestration platform that was initially designed by Google to manage their containers. It has a more complex cluster structure compared to Docker Swarm. Therefore, its setup and installation are not as straightforward. It is composed by master nodes and worker nodes and divided further into pods, namespaces, and config maps. It supports automatic scaling, and it can manage large architectures and complex workloads. Kubernetes is recommended for complex projects that require greater capacity for scalability automation and self-healing (Ganne, 2022).

The Kubernetes cluster will consist of multiple nodes. Typically, it includes a master node and multiple worker nodes. The master node is responsible for managing and controlling the cluster and consists of several key components. One of these components is the API Server, which exposes the Kubernetes API, allowing clients to interact with the cluster and perform operations like deploying applications and services.

Another vital component of the master node is (Wei-guo, Z. et al.,2018), which assigns containers to worker nodes based on resource requirements, availability, and other constraints. It ensures efficient utilization of cluster resources and balanced distribution of containers.

Lastly, two more noteworthy components are the control manager, which manages various types of controllers like the replication controller, node controller, and service controller, and the etcd, which stores the cluster's configuration data, including the desired state of the cluster, such as deployments, services, and replica sets.

In Kubernetes, a pod is the smallest and most basic unit of deployment. It represents a single instance of a running process within the cluster. A pod can contain one or more tightly coupled containers that share resources, network, and storage. Pods are scheduled to run on worker nodes and can communicate with each other using localhost.

Services in Kubernetes provide a stable network endpoint to access a set of pods. They enable load balancing and service discovery within the cluster. A service abstracts the underlying pods and provides a single access point, allowing other pods or services to communicate with the set of pods that the service represents. Services can be exposed internally within the cluster or externally to the outside world.

Deployments in Kubernetes provide a declarative way to manage the lifecycle of pods and ensure application availability. A deployment specifies the desired state of the application, including the number of replicas (pods) to run, container images to use, and update strategies. Deployments manage the creation, scaling, and updating of pods based on the desired state defined in the deployment configuration.

On each worker node, three components are present: Kubelet, container runtime, and Kube proxy. The Kubelet is an agent that runs on each worker node and communicates with the master node. It manages the containers running on the node, ensuring they are started, stopped, and kept healthy according to the instructions received from the master.

The container runtime is responsible for pulling container images, creating, and managing containers, and providing isolation between containers on the node. Kubernetes supports multiple runtimes, such as Docker (Anderson, 2015), CRI-O, and Containerd (Jing et al., 2022).

Lastly, the Kube proxy is a network proxy that runs on each worker node. It handles network communication between services within the cluster, load balances traffic, and provides service discovery. In Fig. 3 we can have an overview of all Kubernetes components.

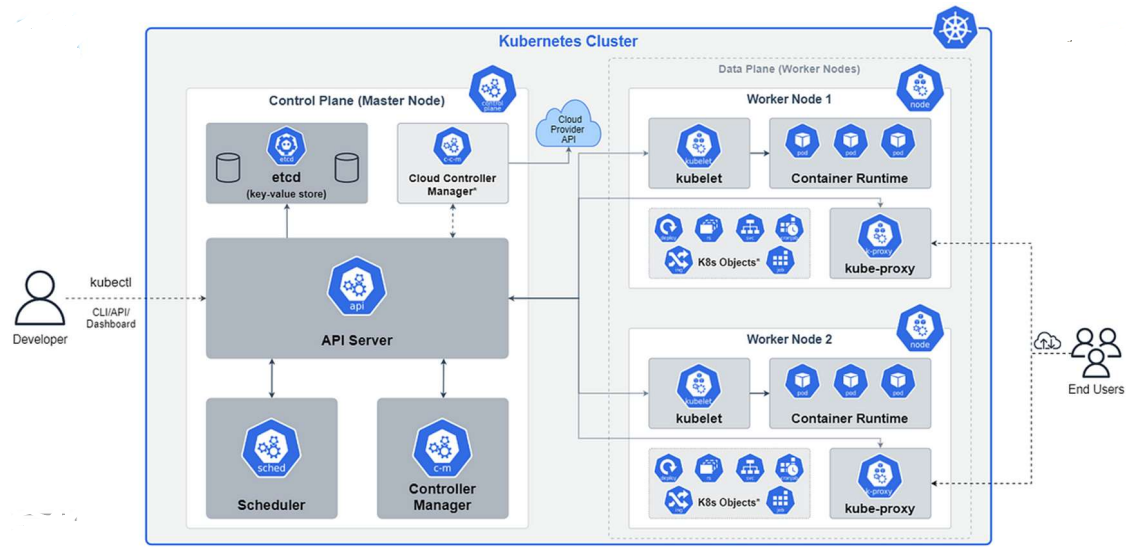

Figure 3 Kubernetes architecture components

### 2.2.3 Other Solutions

- **K3s:** K3s (Böhm & Wirtz, 2021) is a lightweight, certified Kubernetes distribution that is designed for use in resource-constrained environments, such as edge computing and IoT devices. It includes all the core features of Kubernetes, but with some modifications to make it more lightweight and efficient. In general, K3s eliminates the complexity of Kubernetes and provides a lighter and more accessible environment for environments with low resource capacity.

- **OpenShift:** OpenShift (Jakkula, P., n.d.) is a Kubernetes-based container platform developed by Red Hat. It offers features such as container orchestration, build automation, service discovery, and scalability. OpenShift provides additional tools and functionalities on top of Kubernetes, making it a comprehensive platform for managing containerized applications.

- **MicroK8s:** MicroK8s (Böhm & Wirtz, 2021) is another lightweight Kubernetes distribution developed by Canonical, the company behind Ubuntu. It is designed to be a single-node Kubernetes cluster that is easy to install and manage.

- **Cloud-specific Solutions:** Google Kubernetes Engine (GKE) (Shah & Dubaria, 2019), Oracle Container Engine for Kubernetes (OKE) (Jakóbczyk & Jakóbczyk, 2020), Amazon Elastic Container Service (ECS), and AKS are all popular managed container orchestration services (Gunnarsson & Kjeller, 2022). GKE is offered by Google Cloud Platform, providing a scalable and integrated environment for running Kubernetes workloads. OKE, from Oracle Cloud, offers a highly available and secure platform for Kubernetes deployments. Amazon ECS, available on AWS, simplifies container management and scaling. AKS (Shehab & Al-Janabi, 2020),a managed Kubernetes service on Azure, provides robust container orchestration capabilities. Each service has its own unique features and strengths, catering to different needs and preferences in the cloud ecosystem.

## 2.3 Multi-Architecture Nodes

As previously mentioned, in the current state of the CPS and IoT industry, there are several architectures to choose from, each with its own specificities, advantages, and disadvantages. Among the architectures available, there are ARM, x86, and RISC-V.

### 2.3.1 x86

x86 (Aroca & Gonçalves, 2012) processors are a family of CPUs (Central Processing Units) based on the x86 instruction set architecture. This architecture, which is a complex instruction set computer (CISC) architecture, was introduced by Intel in the late 1970s with the release of the Intel 8086 processor. Since then, x86 has become the most widely used CPU architecture in the personal computer industry. It is important to highlight that CISC architectures, including x86, are characterized by many complex and variable-length instructions.

This architecture is known for its compatibility, performance, and widespread support from software developers. It has evolved over the years with various generations and advancements, including the Intel 80286, 80386, 80486, and the popular Intel Pentium series.

In addition to Intel, other companies such as AMD (Advanced Micro Devices) also produce x86 processors. These processors are commonly found in personal computers running operating systems such as Windows, macOS, and Linux. It should be added that macOS late 2020 and beyond use ARM-based architectures. x86 processors are also used in many data centers, powering servers that handle various tasks, including web hosting, cloud computing, and big data processing.

### 2.3.2 ARM

ARM (Amiri-Kordestani & Bourdoucen, 2017) processors, also known as ARM-based processors or simply ARM CPUs, refer to a family of central processing units (CPUs) based on the ARM (Advanced RISC Machines) architecture. ARM processors are designed and developed by ARM Holdings, a company that designs and licenses intellectual property (IP) for various microprocessors and system-on-chip (SoC) designs.

ARM processors are known for their energy efficiency, scalability, and widespread usage in a variety of devices across different industries. They are commonly found in smartphones, tablets, embedded systems, IoT (Internet of Things) devices, automotive applications, and more.

These processors utilize a reduced instruction set computing (RISC) architecture, which focuses on simplicity and efficiency by using a smaller set of instructions that can be executed quickly. This design approach enables ARM processors to deliver high performance while consuming minimal power, making them suitable for battery-powered devices and applications with limited power budgets.

### 2.3.3   RISC-V

RISC-V (Amiri-Kordestani & Bourdoucen, 2017) processors are based on the open-source RISC-V instruction set architecture (ISA). Unlike proprietary architectures like x86 or ARM, RISC-V is freely available for use, modification, and implementation by anyone. It follows the principles of Reduced Instruction Set Computing (RISC), offering a simple and modular design.

One of the main advantages of these processors is its openness. The RISC-V Foundation oversees its development, and the architecture's specifications are open-source, allowing for collaboration and innovation. Its modular nature enables implementers to choose optional extensions, tailoring the design to specific requirements and optimizing performance and power efficiency.

These processors have diverse applications, ranging from embedded systems and IoT devices to smartphones, servers, and high-performance computing. However, it's important to note that RISC-V is still a relatively young architecture compared to x86 and ARM, so it may take time for its adoption to reach the same level. Nevertheless, its openness and customization potential make it an exciting prospect for the future of computer processors.

As mentioned before, several architectures have emerged and gained popularity in various fields. Several systems had to adapt to this trend. This is particularly evident in the case of clusters, where instead of comprising multiple nodes with the same architecture, they leverage the advantages of each architecture and form a multi-architecture cluster.

## 2.4   Compute Unified Device Architecture (CUDA)

The proposed system necessitates the ability to effectively process, classify, and manage vast volumes of data. To accomplish this, it will need a streamlined pipeline capable of harnessing

AI algorithms for classification purposes. To mitigate the heavy computational workload, it is advisable to leverage specific hardware resources, such as the utilization of CUDA cores.

CUDA cores (Ho et al., 2022) are parallel processing units designed to expedite computations on NVIDIA GPUs. These cores represent the individual processing units within a GPU that undertake the actual computational tasks. Their primary responsibility involves executing multiple concurrent threads in parallel, thereby facilitating high-performance computation in various applications such as machine learning, scientific simulations, and data processing.

By using this technology, developers can harness the computational power of GPUs to drastically accelerate calculations that would otherwise consume substantial time on conventional CPU-based systems. This makes CUDA cores particularly well-suited for tasks that can be parallelized, as they can handle many computations simultaneously, leading to faster and more efficient processing.

In short, by employing multiple architectures in the cluster, it will be possible to leverage specific nodes capable of executing more intricate operations utilizing CUDA cores, once they support CUDA, while also incorporating other nodes optimized for efficient management and control tasks. This approach allows for a balanced and optimized utilization of resources across the cluster, catering to diverse computational requirements and enhancing overall performance. By leveraging the strengths of different architectures, the cluster can effectively handle both computationally intensive tasks and management/control operations, resulting in improved efficiency and performance.

When programming with CUDA cores, it is important to be aware of potential support issues that may arise in applications (Cook, 2012). While CUDA cores offer immense parallel processing power and can greatly accelerate computations, certain challenges may arise during development and deployment.

One common issue is compatibility between CUDA cores and different hardware architectures. CUDA cores are specific to NVIDIA GPUs, meaning that applications utilizing CUDA may not run on GPUs from other manufacturers. This can limit the portability of CUDA-based applications and require additional effort to ensure compatibility across different hardware configurations.

## 2.5  Related Projects

This section aims to present similar solutions to ours while highlighting their differences. It is important to note that this section can be divided into two parts: the first part discusses cluster-related projects, while the second part provides a brief overview of data pipeline projects.

### 2.5.1 Cluster-related

In this subsection, there will be two components. Firstly, the presented projects will be within a more academic scope, while in the second part, clustering solutions provided by major cloud providers will be discussed.

#### 2.5.1.1 *ELASTIC*

The ELASTIC architecture (Sousa et al., 2022) aims to enable the efficient and flexible utilization of computing resources in edge and cloud environments. It involves components such as the Distributed Data Analytics Platform, COMPSs (a workflow configuration and coordination framework), DataClay (a data store), Nuvla (a platform for managing applications across the compute continuum), and the Hybrid Fog Computing Platform.

The architecture facilitates the development and deployment of analytic applications, both online and offline, that make use of real-time and historical data from sensors and cameras installed on trams and tram stops. It allows for edge-based execution of real-time applications, such as object detection in the tramway, and cloud-based execution of offline analysis, based on historical data from the edge.

The project also includes a Non-Functional Requirements (NFR) tool component that monitors the execution of workflows and identifies any violations of non-functional requirements, such as missed deadlines. The NFR tool provides online adaptation guidelines to resolve these violations and efficiently manage resources in real-time.

In contrast with our solution that uses Kubernetes, ELASTIC architecture integrates edge and cloud computing mostly with the objective of supporting non-functional requirements, like latency and safety, of smart systems. Kubernetes, on the other hand, is a tool mostly for managing containers and their associated resources.

#### 2.5.1.2 *Edge Cloud Implementation with OpenStack*

The project (Chang, Hari, Mukherjee, & Lakshman, 2014) focuses on implementing an Edge Cloud architecture using the Folsom release of OpenStack, a widely used cloud management platform. The goal of the project is to address the challenges of adapting OpenStack, originally designed for data centers, to effectively support low-cost and low-power platforms in the Edge Cloud.

To achieve this, the project makes several modifications to OpenStack. One significant change is the adoption of lightweight Linux Containers (LXC) as a virtualization primitive instead of hardware virtualization through hypervisors. This allows the Edge Cloud to run on low-power platforms that may not support hardware virtualization.

Despite being similar to our solution, Kubernetes is better suited over OpenStack for our architecture as it is specifically designed to manage containerized applications (Jakkula, P., n.d), being more flexible, adaptable, more tested, and stable. This provides several advantages over traditional virtual machine-based deployments, including faster startup times, better resource utilization, and easier application portability. In terms of application deployment, Edge4CPS provides users with a simple-to-use API and interface that aims to ease and abstract the complexity of operations in the cluster.

### 2.5.1.3  Big Cloud Solutions

In addition to the projects mentioned above, we can find more complex solutions from popular cloud providers. The challenge with major cloud providers is that the majority of their services come with a price tag, making it costly for small businesses to utilize them based on their individual requirements.

- **Amazon Web Services (AWS) Greengrass:** AWS GreenGrass (Ucuz, D., 2020), offered by Amazon Web Services (AWS), extends AWS services to edge devices like industrial machinery, cars, and home automation systems. It enables the execution of AWS Lambda functions, AWS IoT Greengrass connectors, and local compute capabilities on edge devices. GreenGrass supports machine learning inference and provides centralized device management, secure communication, and over-the-air updates for remote devices.

- **Google Cloud IoT Edge:** Google Cloud IoT Edge (Ucuz, D., 2020), part of the Google Cloud IoT platform, offers similar functionalities to AWS GreenGrass. It enables users to deploy and manage workloads on edge devices using lightweight containers. Google Cloud IoT Edge leverages Google Cloud's extensive services and machine learning capabilities, allowing businesses to take advantage of advanced analytics and decision-making at the edge.

- **Microsoft Azure IoT Edge:** Microsoft Azure IoT Edge (Ucuz, D., 2020) is another prominent edge computing solution that empowers organizations to run cloud workloads on edge devices. It offers a container-based environment to deploy and manage modules for local data processing, AI inference, and custom business logic.

Azure IoT Edge integrates seamlessly with Azure services, providing a comprehensive ecosystem for edge computing, data analytics, and device management.

### 2.5.2  Data Pipeline Solutions

Regarding the data processing aspect, data pipelines, one of the components provided in Edge4CPS, are an important consideration, once it ensures that data is properly managed, transformed and made ready for analysis in any given context. Related to this there are several projects that already implemented them. For example, in (Garan, Tidriri, & Kovalenko, 2022), the authors developed a data-centric machine learning methodology for predictive maintenance of wind turbines, which incorporated the use of a data pipeline. The pipeline processed a dataset of sensor readings from wind turbines and transformed the data to prepare it for analysis by machine learning algorithms, which were then used to predict potential failures and identify maintenance needs. Moreover, the Smart-PDM (Chaves et al., 2022) project also developed a system that uses a data pipeline to process and classify data from home appliances for predictive maintenance. The main problem with these solutions is related with its scalability, since all the applications work on the same machine. Smart-PDM solution already offer some predefined and limited scalability solutions. To address this issue, our goal is to cluster the data pipeline, separating their applications and services through multiple nodes.

# 3  Technical Description

This chapter presents the requirements and preliminary analysis of the project, aiming to outline the specific functionalities, features, and constraints that need to be addressed to fulfil the project's goals.

## 3.1  Requirements Engineering

Requirements Engineering (Jitnah et al., 1995) is a systematic process within the field of software engineering that focuses on understanding and documenting the needs and characteristics of the software system to define its scope, functionality, and behaviour. This process involves understanding the user's requirements, business goals, and operational constraints to ensure that the software solution meets the desired objectives.

### 3.1.1  Functional Requirements

Functional requirements (FR) ("Functional vs. Non-functional Requirements." QRA Corp, 2023) encapsulate the precise behaviours, tasks, and operations that a software system is mandated to undertake in order to satisfy the exigencies of its users. These requirements elucidate the anticipated functionality of the system, delineating its expected input, processing, and output mechanisms. More specifically, they serve as a comprehensive compendium of desired features, actions, and interactions that end-users can seamlessly engage in during their interaction with the software.

In this case, following a comprehensive analysis of all project specifications and extensive discussions with the project members, the functional requirements of the project are as follows:

**FR 1.** The system should be running on a clustered environment.

**FR 2.** The system must run users' applications and services on a containerized environment.

**FR 3.** The system must provide a web interface for the user to deploy applications to run in the system.

The User must fill out the form with the specific attributes of the container to be able to deploy:

- Container Name.
- Ports to be exposed inside the cluster.

- Image Name to be deployed.
- Define CPU limitations.
- Define Memory limitations.

**FR 4.** The system should be able to handle multiple users deploying applications and services to run on the system.

**FR 5.** The users must be authenticated before deploying applications and services.

To authenticate the user must fill one form with two fields:

- API Key
- Folder Id (Serves as an identifier)

**FR 6.** The system must provide the user with an API capable of interacting with the system.

**FR 7.** The system must be capable of supporting a messaging protocol middleware.

**FR 8.** The middleware implemented should support multiple protocols.

**FR 9.** The system must have a pipeline running for data processing, classification and visualization that demonstrates the system's environment.

### 3.1.2  Non-Functional Requirements

Non-functional (NFR) (Glinz, 2007) requirements are essential aspects of software that go beyond its functional capabilities. They encompass implicit or expected characteristics, quality attributes, constraints, and restrictions of the system. Taking into consideration the nature of these requirements, the primary non-functional requirements for the project can be summarized as follows:

#### 3.1.2.1  *Usability*

Usability ("Usability." Computer Science Wiki, 2023) is a measure of how user-friendly and easy to use a software system or application is. It focuses on ensuring that the interface and functionality of the software are intuitive and efficient, allowing users to achieve their goals effectively.

**NFR 1.** The web interface responsible for the deployment of applications and services should be designed to be simple and intuitive. By providing an intuitive interface, it minimizes user errors during the deployment process.

**NFR 2.** The pipeline inside the cluster will include a component dedicated to data visualization. It is essential for this component to be intuitive and visually appealing, enabling users to easily comprehend the information being presented to them.

### 3.1.2.2 Performance

Performance ("Computer Performance Evaluation: Definition, Challenges & Parameters." Study.com, 2023) relates to the speed, efficiency, and responsiveness of a software system.

**NFR 3.** The messaging protocol middleware implemented in the system must efficiently translate to others messaging protocols.

**NFR 4.** To ensure efficient classification in the data pipeline, it is crucial for the process to operate without delays. To achieve this, the process will be parallelized.

### 3.1.2.3 Scalability

Scalability ("Scalability in Software Development."., n.d.) refers to a software system's ability to handle an increasing workload or accommodate growing user demands without a significant drop in performance. It involves designing the software in a way that allows it to scale up or down seamlessly. Scalability can be achieved through various techniques, such as horizontal scaling (adding more hardware resources) or vertical scaling (improving the capacity of existing resources).

**NFR 5.** The system should have the capability to integrate new nodes quickly and seamlessly. If the existing resources are insufficient to handle the current load. This flexibility enables efficient scaling of the system to meet increased demands.

### 3.1.2.4 Interoperability

Interoperability ("Interoperability.", n.d.) refers to the ability of a software system to communicate and work effectively with other systems or components. It involves ensuring that different software applications or modules can exchange data and utilize each other's functionality seamlessly.

**NFR 6.** The messaging protocol middleware should provide an abstraction layer that facilitates seamless communication between different protocols, granting users the ability to interact with others using their preferred protocol.

### 3.1.2.5 Portability

Portability ("Portability." TechTarget, n.d.) refers to the ease with which software can be transferred or adapted to different computing environments or platforms. It involves designing the software in a way that minimizes dependencies on specific hardware, operating systems, or software libraries. Portable software can run on various platforms without significant modifications, providing flexibility and reducing development and maintenance costs.

**NFR 7.** This system must be capable to implement on different types of OS and CPU architectures.

**NFR 8.** The system should be designed to be compatible with and capable of being implemented on various cluster technologies.

## 3.2 Actors

In the Edge4CPS system, there will be two distinct types of actors that will play vital roles in facilitating and managing all interactions within the system: users and system administrators.

- **Users:** The user will be the actor who will have access to both the Edge4CPS API and web interface, enabling them to deploy applications and services, as well as check which ports their services have been exposed on. They will also be able to utilize cluster-provided services such as middleware and messaging protocol brokers. However, authentication will be required for them to access and benefit from the system.

- **System Administrator:** The system administrator is the actor who will manage and be in closer contact with the critical components of the cluster, thus having access to the entire system. Effectively, the system administrator will have access to the Kubernetes dashboard, as well as physical and remote access to the cluster and all the nodes within it. Finally, they will also have the ability to use the implemented data pipeline and test the system using the pipeline.

## 3.3 User Stories

User stories are a technique used in agile software development to capture and describe requirements from the perspective of end users or stakeholders. They are concise, informal descriptions of a specific desired functionality or feature of a software ("User Stories." Mountain Goat Software, n.d.).

In this project, after a discussion with the advisor and project members, we arrived at the following user stories:

Table 2 User Stories & Acceptance Tests

| User Stories | Acceptance Test |
| --- | --- |
| *As a user I want to access the Edge4CPS Web Interface* | The user must access the platform's web address and login |
| *As a user I want to be able to deploy applications using the web interface* | The user must fill out the required forms: container Name, ports to be exposed inside the cluster, image name, CPU limitations, memory limitations. |
| *As a user I want to be able to deploy applications using the Edge4CPS API* | The user must fill out the required parameters, when making the HTTP/HTTPS request: container Name, ports to be exposed inside the cluster, image name, CPU limitations, memory limitations. |
| *As a user I want my service or application to run on a containerized environment* | The system where the user will deploy the application or service must have a containerized environment. |
| *As a user I want to be able to access the messaging protocol middleware* | The user must be given access to the middleware through its address. |
| *As a user I want to know what services are exposed to what port* | The user must have networking information about their services and applications |
| *As a system administrator I want my system to be scalable* | The system must be implemented as a cluster of networked nodes |
| *As a system administrator I want to be able to make operations inside the cluster* | The administrator must have physical access to the cluster console (e.g., to the Kubernets dashboard) |
| *As a system administrator I want to use the data pipeline implemented inside the system* | The administrator must have access to the data pipeline, which includes the classification, data processing, and visualization components. |

## 3.4  Analysis

Domain analysis (Tracz, W., 1992) is a systematic process used in software development to gather, document, and structure information that is specific to a particular domain or field. Its main objective is to create a domain model that encapsulates the essential concepts, behaviours, and data related to the domain, making this information reusable for future software systems.

During domain analysis, a software engineer aims to learn and understand the business or problem domain in which the software will be used. They identify and capture relevant information about the domain, such as its core concepts, processes, rules, and logic. This information is then organized and represented in a domain model.

### 3.4.1  Domain Model

Based on the previous requirements discussed, the Fig. 4 presents the domain model developed.

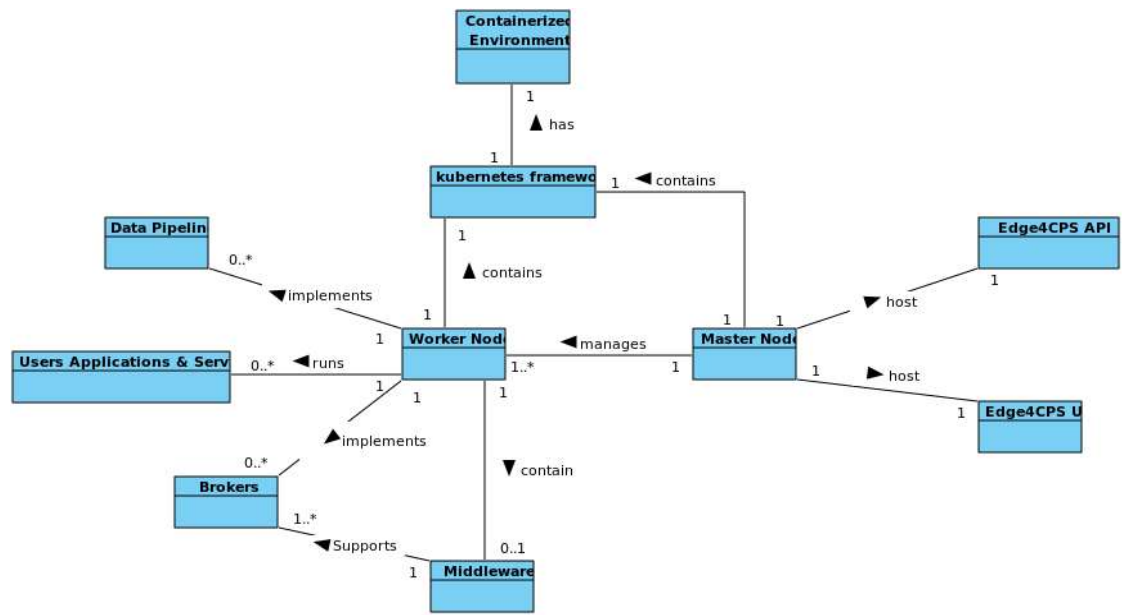

Figure 4 Domain Model Edge4Cps

**Master Node:**

The Master Node class represents the node that is responsible for managing and distributing tasks to all the other nodes in the system.

- **Edge4CPS API:** The Edge4CPS API class represents the API responsible for interacting with the cluster console. It is responsible for deploying applications and services and checking for exposed ports. This API provides an endpoint for users to deploy applications within the cluster.

- **Edge4CPS UI:** The Edge4CPS UI class represents the user interface that will be exposed to users. It is responsible for abstracting the Edge4CPS API and providing users with a graphical and simple way to deploy applications and services, as well as check the exposed ports of their applications.

**Worker Node:** The Worker Node class, symbolized in the domain model, represents a crucial component within the system architecture. It encompasses all the individual nodes that are dedicated to executing tasks delegated by the master node.

- **Data Pipeline:** The Data Pipeline class, represented in the domain model, embodies the core functionality related to data processing, classification, and visualization within the system. It serves as a pivotal component that orchestrates these tasks, and it operates as a service running on a dedicated worker node.
- **Users Applications & Services:** This class, depicted in the domain model, encompasses all the services and applications that are deployed and run within the worker node. These services and applications are deployed by users utilizing either the Web Interface or the API.
- **Brokers:** The Brokers class, depicted in the domain model, represents the messaging protocols that operate within the system, specifically within the worker node. These protocols facilitate communication between devices and applications within the system.
- **Middleware:** The Middleware class represents a service that can translate multiple messaging protocols. It interacts with the brokers to handle the translation between protocols. This class is running on the worker node.

**Kubernetes Framework:** The Kubernetes class represents the component that runs within both the master node and the worker node. This framework enables the implementation of the system in a clustered manner.

- **Containerized environment:** The Containerized Environment class represents the primary environment where most applications and services will run within the cluster. It virtualizes the applications, providing a separate and isolated runtime environment for each. This environment is essential for the proper functioning of the cluster framework.

# 4  Design

Software design encompasses a systematic and iterative process that involves analysing, specifying, and transforming user requirements into a well-structured and coherent design ("Software Design Basics.", n.d.).

After thoroughly assessing both the functional and non-functional requirements of the project and conducting a comprehensive analysis, this chapter aims to provide a detailed overview of the decisions that have been made, including a thorough exploration of the Edge4CPS architecture.

## 4.1  Architectural Design

This sub-chapter will describe the system architecture design by applying the C4 model. (Vázquez-Ingelmo, García-Holgado, & García-Peñalvo, 2020).

### 4.1.1  Level 1

This section provides an overview of the system's structure, including its core components in the Logical View and the system's use cases in the Scenarios view.

#### 4.1.1.1  Logical View

The Level 1 Logical View in Fig. 5 shows an overview of the developed system architecture and components.

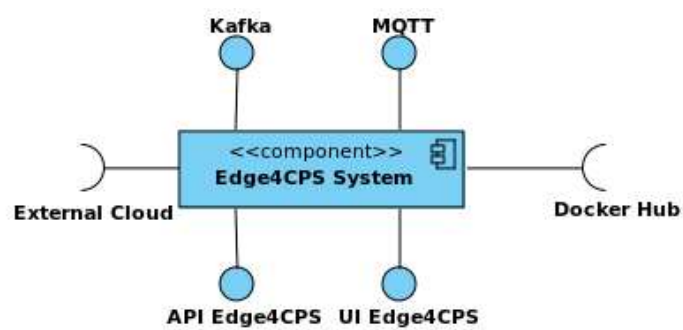

Figure 5 Level 1 Logical View

The system components and endpoints presented in the Level 1 Logical View are:

- **Edge4CPS System** – This is the main component, and it is where the cluster applications and services will be contained, along with the messaging protocol middleware, data processing pipeline, classification, and visualization. The Edge4CPS

System component consumes two APIs, delivers one API, and has three endpoints: one for the Interface and the other two related to broker endpoints.

- **Brokers endpoints (Kafka and MQTT)** – These two endpoints provide connection to devices that want to communicate with the brokers inside the system.
- **UI Edge4CPS** – Delivers a graphical User interface to the End User.
- **API Edge4CPS** – Provides an API endpoint to deploy applications and services inside the cluster.
- **Docker Hub** – The Docker Hub connection is used when the containerized environment needs to retrieve the user's image from the external Docker Hub repository.
- **External Cloud** – This connection to this API may or may not exist. If it exists, it functions as an extension of the system to provide external deployment on a remote cloud.

### 4.1.1.2  Scenarios View

The Scenarios View is a perspective that focuses on capturing and analysing the different scenarios or use cases ("Use Cases.", n.d.) that the software system is designed to support.

Given the functional requirements (section 3.1.1), the following use cases were created:

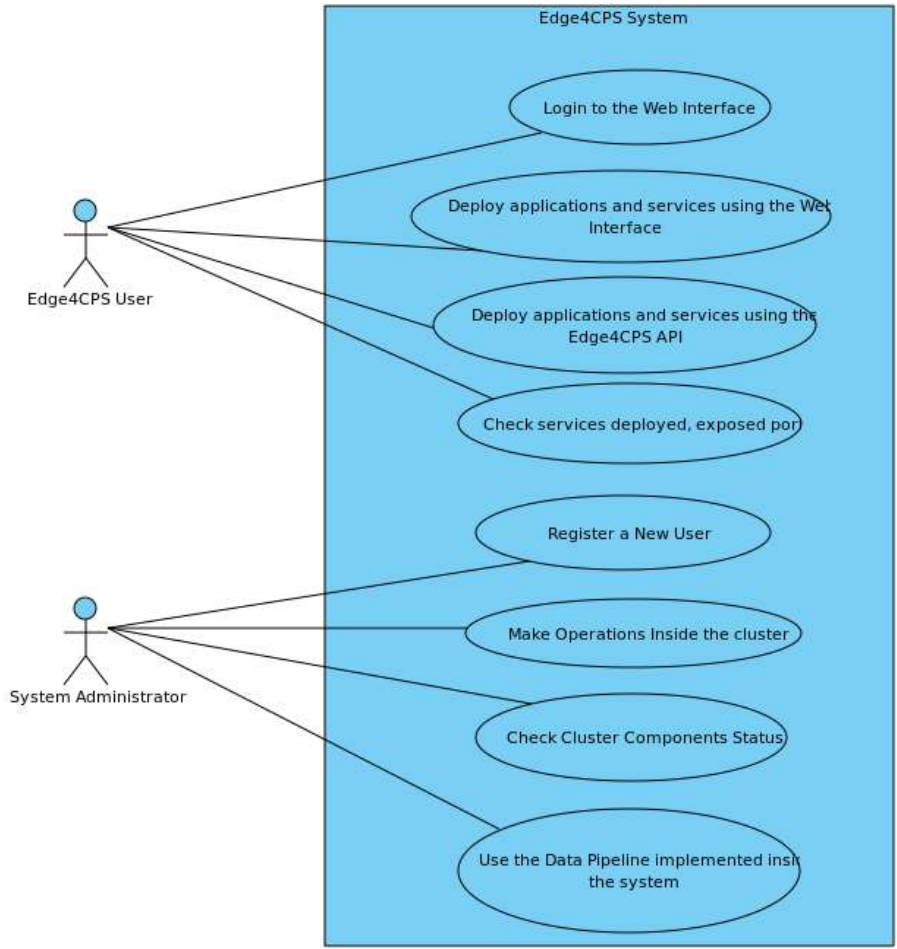

Figure 6 Use Cases Edge4CPS

### 4.1.2 Level 2

Level 2 provides additional insights into the system design by breaking it down into interconnected modules or components, while explicitly outlining the packaging and processes involved in the system implementation.

#### 4.1.2.1 Logical View

The Level 2 Logical View (Fig. 7) zooms into the Edge4CPS system component illustrated in the Level 1 Logical View, showing with increased granularity, it presents a more detailed depiction of the containers that constitute the designed solution.

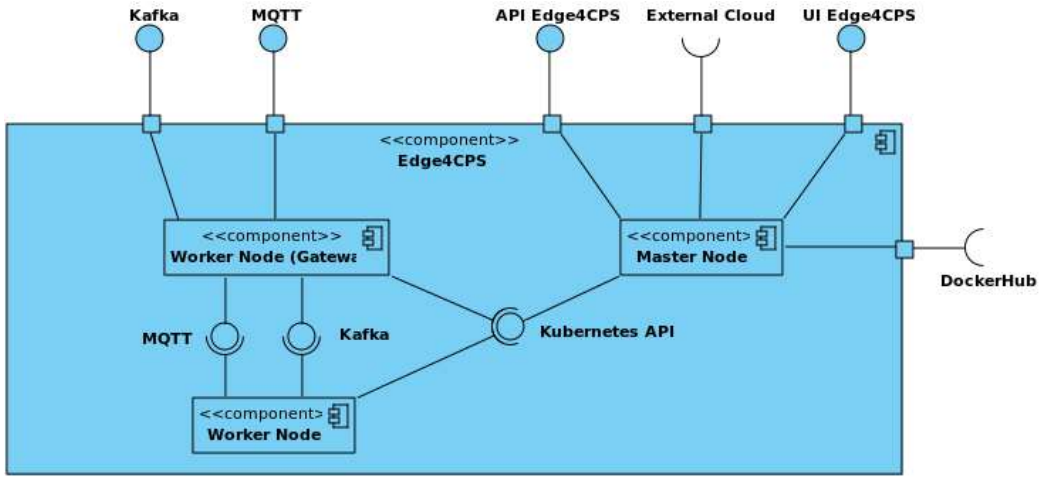

Figure 7 Level 2 Logical View

In the level 2 logical view there are three components:

- **Master Node** – The master node component will assume the crucial role of overseeing the worker nodes, utilizing the Kubernetes API endpoint. Furthermore, it will encompass both the Edge4CPS API and the Web Interface. The containerized environment inside the master node will have a connection to the Docker Hub to be able to gather users' applications to deploy.

- **Worker Node** – The worker node leverages the Kubernetes API to receive task assignments from the master node, specifically related to deploying applications and services. Additionally, it encompasses a data pipeline responsible for data processing, classification, and visualization. To acquire the data, the worker node establishes connections to brokers facilitated by the gateway.

- **Worker Node (Gateway)** – The gateway assumes the role of a worker node, considering that the master node will deploy the two broker services, Kafka ("What is Apache Kafka?", n.d.) and MQTT ("What is MQTT and How Does it Work? A Beginner's Guide.", n.d.). Subsequently, these brokers will be exposed, providing an endpoint through which other nodes can establish communication.

#### 4.1.2.2 Implementation View

The Level 2 Implementation View displays the distribution of packages within the Edge4CPS system. Fig. 8 provides a visual representation of the three key packages along with their respective dependencies.

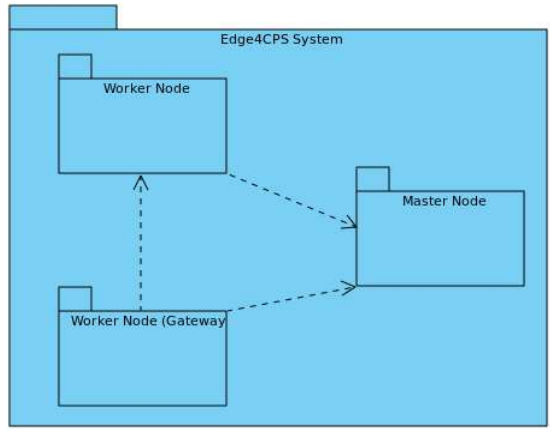

Figure 8 Level 2 Implementation View

In the following Fig. 9, it will be possible to observe the mapping of the two views, the logical view level two, and the implementation view also at level two. With the purpose of interconnecting the packages of the implementation view with the components of the logical view.

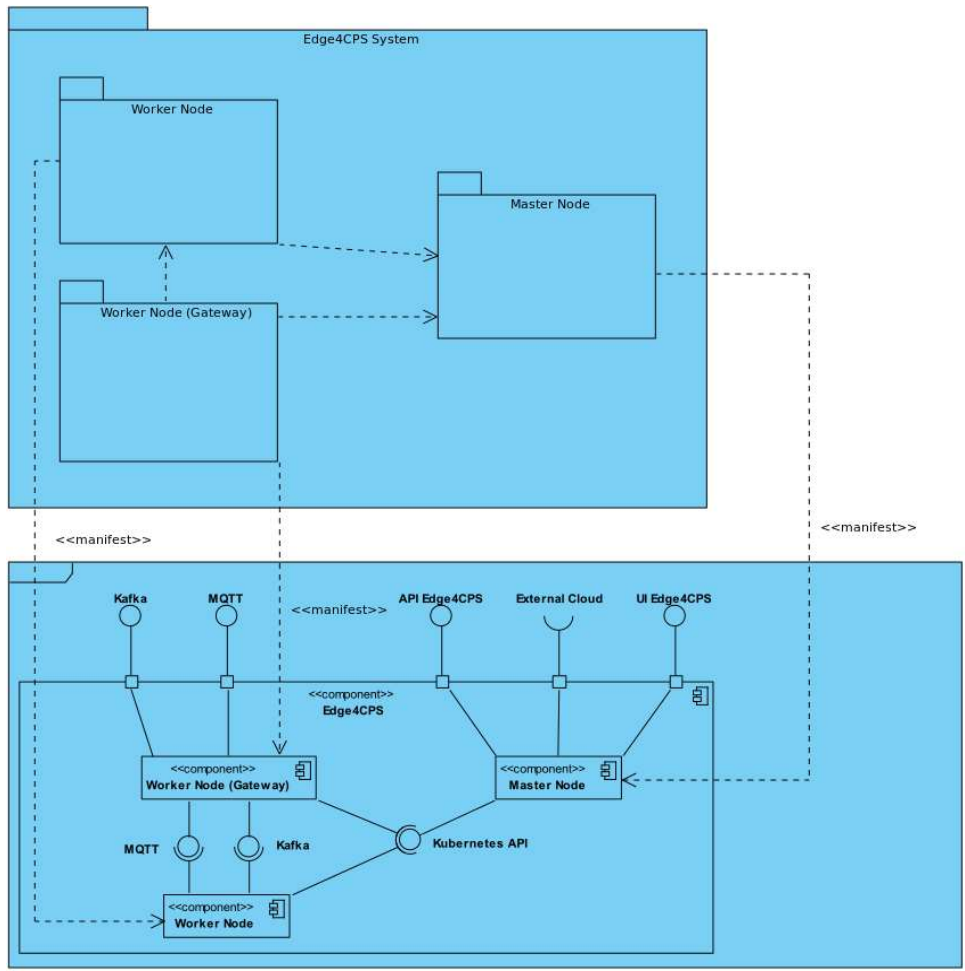

Figure 9 Level 2 Implementation View vs Logical View

### *4.1.2.3 Process View*

The Process View at Level 2 provides a depiction of each system Use Case through the utilization of UML sequence diagram notation ("Unified Modeling Language (UML) - Sequence Diagrams.", n.d.). Furthermore, the subsequent sections will outline detailed information regarding each use case, including its description, associated actor, pre- and post-conditions, and the required data.

## Use Case 1

Table 3 Use Case 1: Login to the Web Interface

| *Use Case 1* | *Login to the Web Interface* |
|---|---|
| *Description* | The user intends to login to the interface to be able to use it. |
| *Actor(s)* | Edge4CPS User |
| *Preconditions* | 1. Have access to the Edge4CPS web interface |
| *Postconditions* | 1. The user is logged in the web interface and can use it. |
| *Necessary data* | • API key<br>• Folder id |

*Base Flow events*

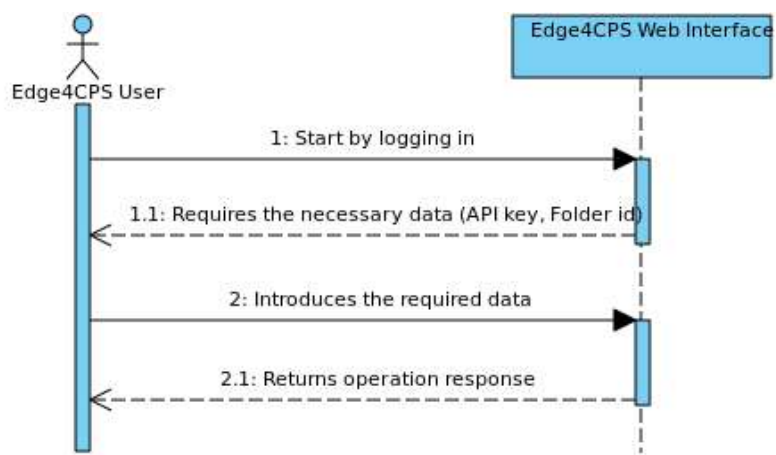

Figure 10 SSD UC1: Login User

## Use Case 2

Table 4 Use Case 2: Deploy applications and services using the Web Interface

| Use Case 2 | Deploy applications and services using the web interface |
|---|---|
| Description | The user intends to deploy applications inside the system using the web interface |
| Actor(s) | Edge4CPS User |
| Preconditions | 1. Have access to the Edge4CPS web interface.<br>2. Logged in to the Edge4CPS web interface. |
| Postconditions | 1. The user sent a deployment order to the API to run the application or service inside the cluster. |
| Necessary data | • Container name.<br>• Ports to be exposed.<br>• Image name.<br>• CPU limitations.<br>• Memory limitations. |

Base Flow events

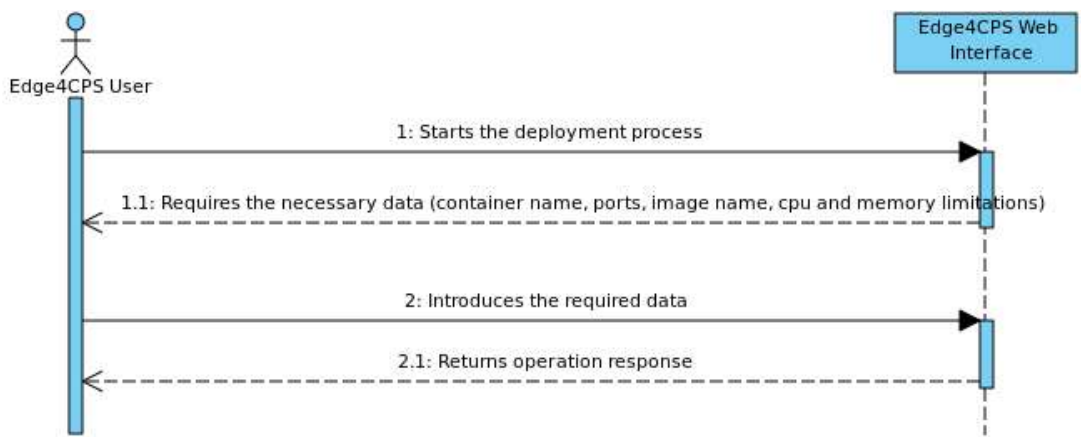

Figure 11 SSD US2: Deploy applications using the Web Interface

## Use Case 3

Table 5 Use Case 3: Deploy applications and services using the Edge4CPS API

| *Use Case 3* | *Deploy applications and services using the Edge4CPS API* |
|---|---|
| *Description* | The user intends to deploy applications inside the system using the Edge4CPS API |
| *Actor(s)* | Edge4CPS User |
| *Preconditions* | 1. Have access to the Edge4CPS API.<br>2. Have the credentials to perform the request. |
| *Postconditions* | 3. The user sent a deployment order to the master node to execute the deployment. |
| *Necessary data* | • API key<br>• Folder id<br>• Container name.<br>• Ports to be exposed.<br>• Image name.<br>• CPU limitations.<br>• Memory limitations. |

*Base Flow events*

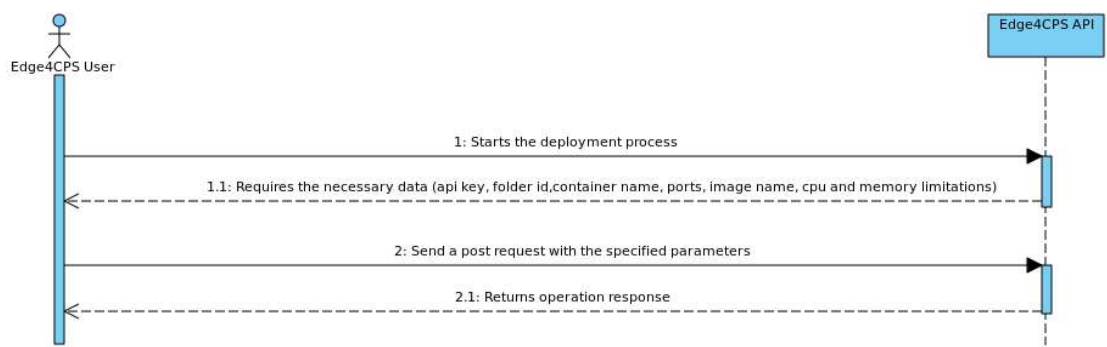

Figure 12 SSD US2: Deploy applications using the Edge4CPS API

## Use Case 4

Table 6 Use Case 4: Check Services Deployed Exposed Ports

| *Use Case 4* | *Check Services deployed, exposed ports* |
|---|---|
| *Description* | The user intends to check what ports are exposed in their applications and services. |
| *Actor(s)* | Edge4CPS User |
| *Preconditions* | 1. Have access to the Edge4CPS web interface. 2. Have access to the Kubernetes web interface. 3. Login to the Kubernetes web interface |
| *Postconditions* | 1. The user can check inside the Kubernetes dashboard what ports are exposed in his applications. |
| *Necessary data* | • Kubernetes Token |

*Base Flow events*

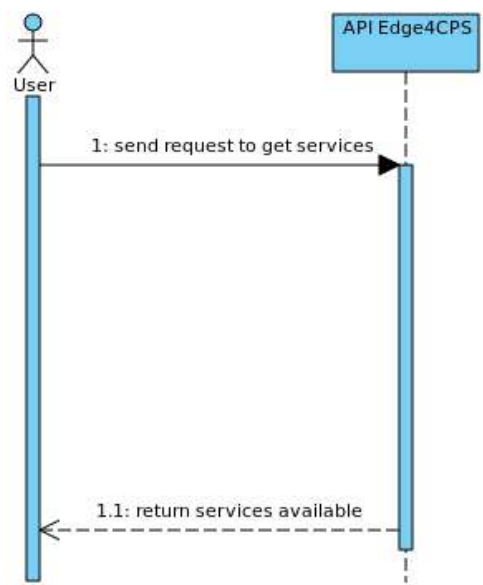

Figure 13 SSD UC4: Check applications exposed ports

## Use Case 5

Table 7 Use Case 5: Register a New User

| Use Case 5 | Register a new User |
|---|---|
| Description | The administrator intends to register a new user, enabling the user to utilize the Edge4CPS API and web interface. |
| Actor(s) | System administrator |
| Preconditions | 1. Have access to the Edge4CPS API.<br>2. Have the credentials to perform the request. |
| Postconditions | 1. The administrator successfully registers and grants permission to the user to utilize the Edge4CPS API and web interface. |
| Necessary data | • Edge4CPS API key<br>• New Folder id<br>• User API key |

*Base Flow events*

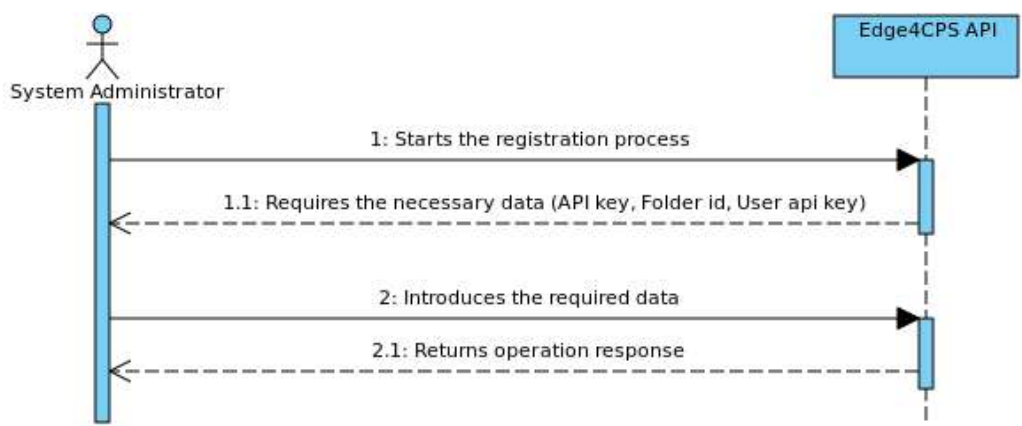

Figure 14 SSD UC5: Register a New User

## Use Case 6

Table 8 Use Case 6: Make Operations Inside the Cluster

| Use Case 6 | Make Operations Inside the Cluster |
|---|---|
| Description | The administrator intends to perform operations inside the master node of the cluster by SSH. |
| Actor(s) | System administrator |
| Preconditions | 1. Have access to the master node. 2. Have the credentials to authenticate via SSH to the master node. |
| Postconditions | 1. The administrator successfully interacts via console with the cluster master node through SSH. |
| Necessary data | • SSH username • SSH password |

Base Flow events

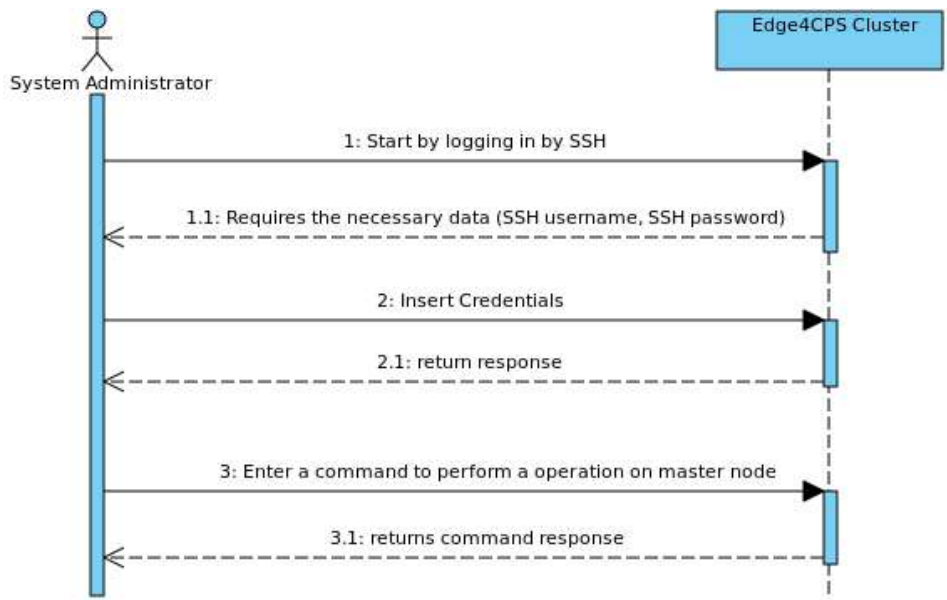

Figure 15 SSD UC6: Make operations inside The Cluster

## Use Case 7

Table 9 Use Case 7: Check Cluster Components Status

| Use Case 7 | Check Cluster Components Status |
|---|---|
| Description | The system administrator intends to check the health status of the cluster's essential applications and services through the Kubernetes dashboard. |
| Actor(s) | System Administrator |
| Preconditions | 1. Have access to the Edge4CPS web interface.<br>2. Have access to the Kubernetes web interface.<br>3. Login to the Kubernetes web interface |
| Postconditions | 1. The system administrator can check inside the Kubernetes dashboard all the status and reports from essential applications and services running inside the cluster. |
| Necessary data | • Kubernetes Token |

Base Flow events

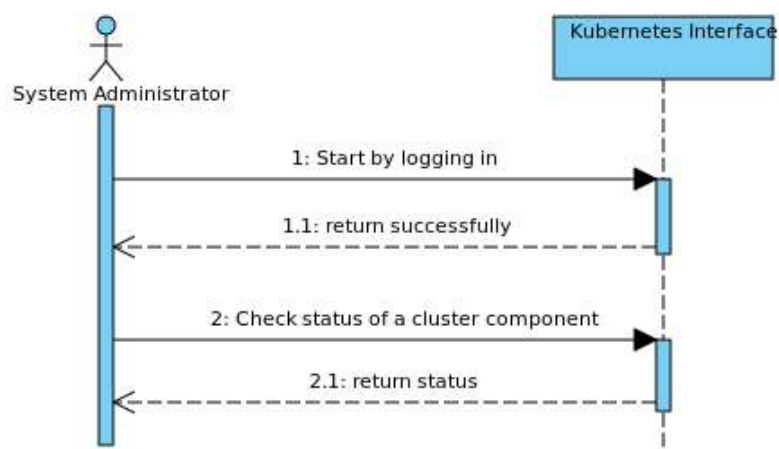

Figure 16 SSD UC7: Check cluster components status

## Use Case 8

Table 10 Use Case 8: Use the data pipeline implemented Inside The System

| Use Case 8 | Use the data pipeline implemented inside the system |
|---|---|
| Description | As a system administrator intends to use the data pipeline implemented inside the system |
| Actor(s) | System Administrator |
| Preconditions | 1. Have access to the data pipeline |
| Postconditions | 1. The system administrator can access the classification, data processing and visualization components. |
| Necessary data | • Select which component want to see |

Base Flow events

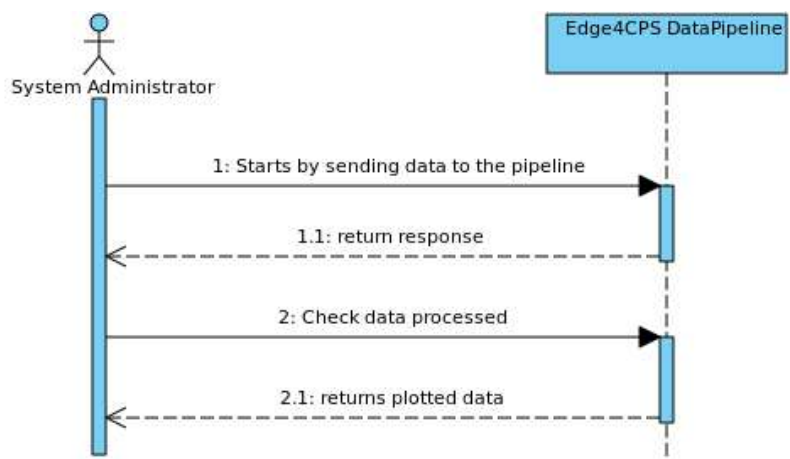

Figure 17 SSD UC8: Visualize classified and real-time data from the Data Pipeline

### 4.1.3   Level 3

In this section, we will delve into a detailed examination of each container through a Level 3 zoom-in. This will provide a comprehensive depiction of a series of interconnected components that seamlessly integrate within the Logical View, the correspondent deployment view, and the use case processes in the Process View.

### 4.1.3.1  Logical View

At this level, the Logical View will be divided into three distinct components, as a result of zooming in on the system. This closer examination reveals the presence of the Gateway, the Master Node, and the Worker Node. Each of these components plays a crucial role in the overall functioning of the system.

## Master Node

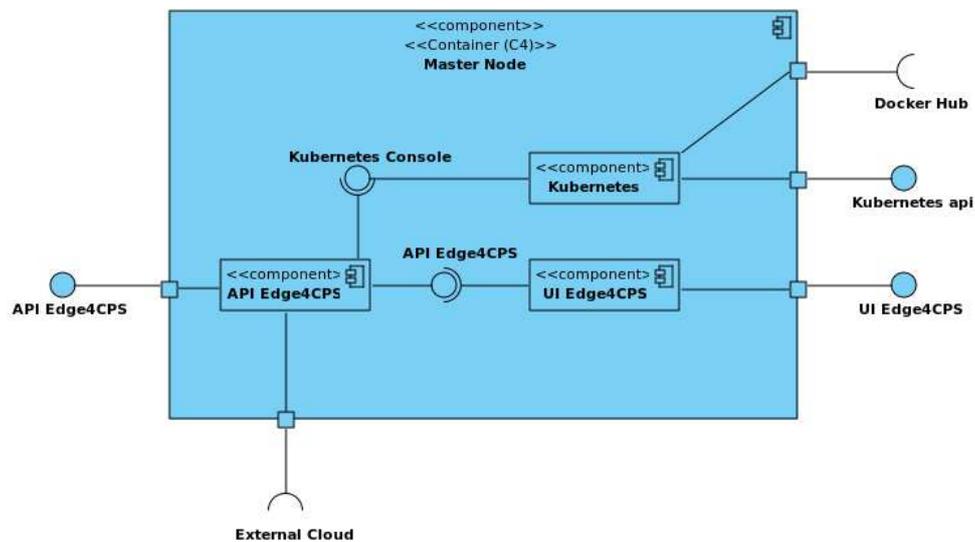

Figure 18 Level 3 Logical View (Master Node)

- **Kubernetes** – The Kubernetes component within the Master Node assumes the vital responsibility of overseeing and coordinating all other nodes, as well as assigning tasks to its workers. It exposes the Kubernetes API, enabling other nodes to join the cluster and receive tasks from the master. Furthermore, within this node, an additional endpoint is made available, the Kubernetes console. This console provides access to an API that facilitates direct interaction with the cluster. All the applications and services running inside the Kubernetes are deployed inside a containerized environment. The Docker Hub ("What is Docker Hub?" GeeksforGeeks, n.d.) connection originates from the containerized environment, specifically the Docker environment. This connection enables the Docker environment to fetch the application that will be deployed in the system from the Docker Hub repository, assuming the image is stored there.

- **API Edge4CPS** – The Edge4CPS API integrates with the Kubernetes framework through its endpoint, enabling seamless interaction with the cluster for performing operations. It exposes an endpoint that allows users to request the deployment of services and applications within the worker nodes. The primary objective of the Edge4CPS API is to act as an intermediary between the user and the cluster, streamlining the deployment process and making it more accessible.

- **UI Edge4CPS** – The Edge4CPS UI has been implemented within the master node, utilizing the Edge4CPS API endpoint. This design allows for abstraction of the request-sending process from the user's perspective, providing a user-friendly graphical interface. The UI is responsible for facilitating the request to deploy applications and services to the API, offering a seamless experience for users to interact with the system.

## Worker Node

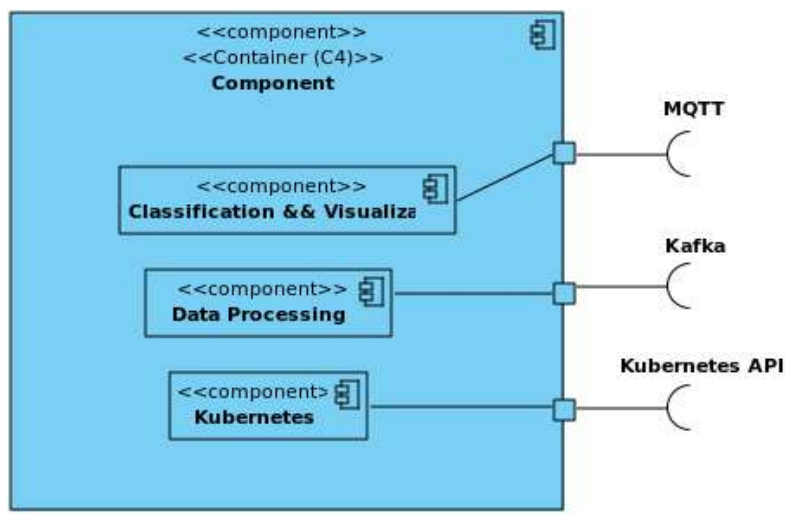

Figure 19 Level 3 Logical View (Worker Node)

- **Kubernetes** – The Kubernetes component running within the Gateway plays a vital role in executing tasks received from the master node, effectively serving as a worker node. To establish seamless communication with the master node, it subscribes to the Kubernetes API endpoint running within the master node.

- **Data Processing** – The Data processing component is an integral part of the data pipeline service implemented within the cluster. Its primary responsibility is to collect data from the MQTT broker endpoint provided by the gateway. Subsequently, the collected data is processed before being sent back to the broker. The middleware, provided within the gateway, then translates the data into Kafka format.

- **Classification and Visualization** – The classification and visualization components play a crucial role in the subsequent step of the system. They retrieve the data translated from the middleware to Kafka using the endpoint provided by the gateway. The data is then subjected to classification using AI algorithms. Once classified, both the processed data and the classified data are presented in a graphical application. This graphical interface provides the system administrator with a user-friendly environment to gather comprehensive information about the data.

## Worker Node (Gateway)

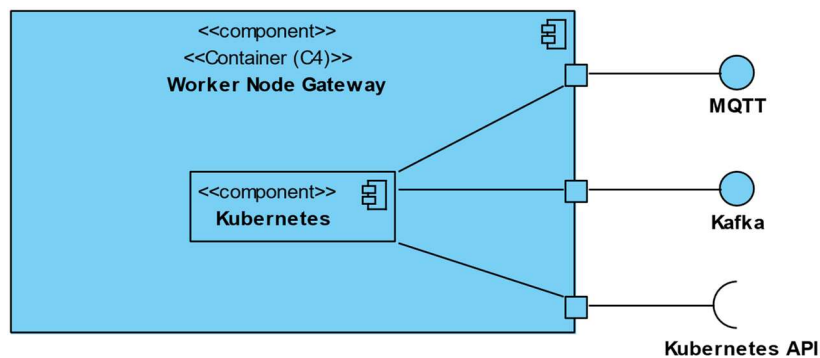

Figure 20 Level 3 Logical View (Gateway)

- **Kubernetes** – In addition to functioning as a worker node, the Gateway node holds unique responsibilities within the system. Within the containerized environment of the Kubernetes framework, this node operates two messaging protocol brokers alongside a middleware. These brokers expose their endpoints, enabling other applications to utilize them and facilitating seamless communication across the system. The middleware plays a vital role in translating between different messaging protocols, providing users with the flexibility to dynamically choose and utilize the most suitable protocol for their needs.

### 4.1.3.2   Deployment Diagram

In this level, a deployment diagram ("Deployment Diagram Tutorial.", n.d.) has been chosen to represent the architecture. The deployment diagram provides a visual representation of how the software components and their interactions are mapped onto the underlying hardware or execution environment. This diagram helps in understanding the physical deployment architecture and how the software will be installed, executed, and integrated within the system. It offers valuable insights into the distribution of components and their relationships with the hardware. Additionally, a deployment diagram, Fig. 21, showcases how the components interact at the protocol level, providing an understanding of the communication patterns and protocols used within the overall deployment structure.

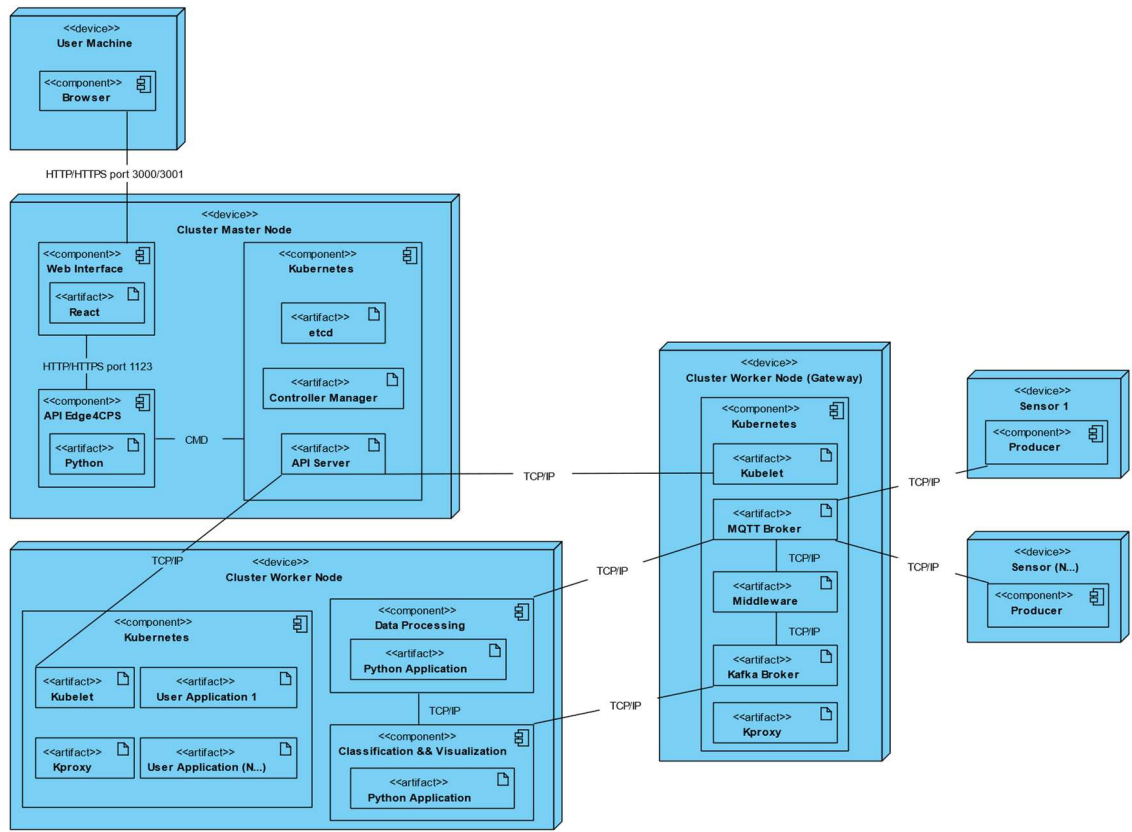

Figure 21 Edge4CPS deployment diagram

**Cluster Master Node:** As previously discussed, the master node will serve as the head of the cluster. Effectively, the components within it are responsible for ensuring the proper functioning of the entire system. This node will be running on a machine with sufficient resources to ensure the smooth operation of all system applications. If it goes down, a part of the system will cease to function.

- **Kubernetes:** This component is the framework responsible for running distributed systems resiliently. Since it serves as the master node, this component will be responsible for communicating and distributing tasks among the worker nodes, as previously mentioned. For communication, the framework defaults to using the TCP/IP ("Web, Internet, TCP/IP, and HTTP concepts.", 2023) protocol on port 6443, based on conventions within the Kubernetes ecosystem. In production environments, it will utilize port 443.

- **Web Interface:** The Web Interface component will provide an external user with a way to interact with the cluster, as previously mentioned. It offers the capability to deploy applications on the cluster and will have an endpoint on port 3000/3001, depending on whether it is HTTP or HTTPS ("Web, Internet, TCP/IP, and HTTP concepts.", 2023). Additionally, it will communicate with the Edge4CPS API endpoint on port 1123 to make deployment requests.

- **API Edge4CPS:** The Edge4CPS API has an endpoint on port 1123, which is used by the Web Interface. It can also be utilized by other developers who wish to implement a different interface. The API is hosted on the same device where the Kubernetes framework of the master node resides, as the API utilizes the Kubernetes console to execute commands on the cluster.

**Cluster Worker Node:** This node serves as a worker device to the master node, providing computing power to the cluster. This node has the capacity to support and run various applications from users. It shares with other nodes the resources to

- **Kubernetes:** This component, similar to the one present in the master node, differ in its role as a worker node. Effectively, in this role, it is responsible for executing tasks sent from the master node. To communicate with the master node, the worker node utilizes TCP/IP.

- **Data Processing:** The data processing component will be part of the data pipeline. Its task will be to receive data through the MQTT messaging protocol, which operates over TCP/IP, and subsequently clean and process the data. Finally, after this processing, the data will be sent back via MQTT to be used by the classification component, which will be discussed later.

- **Classification:** Also related to the data pipeline of the system, there will be the classification and visualization component. Once the data arrives from the Kafka broker's endpoint via the TCP/IP protocol, this component will utilize the processed data from the previous stage to apply AI algorithms for classification, as discussed earlier.

- **Visualization:** Finally, the processed and classified data will be displayed in a visualization interface. It is worth noting that the pipeline resides within the worker nodes again, given their higher processing capacity demands.

**Cluster Worker Node (Gateway):** Regarding the gateway node, it will also function as a worker node, having the same connections with the master node. However, in this case, the applications running on this node will serve to provide communication protocols.

- **Kubernetes:** This component performs the same operations as the component with the same name on the worker node. However, as a novelty, instead of having user applications running in a containerized environment, it will contain messaging protocols, similar to a middleware service. The middleware service will utilize both Kafka and MQTT brokers to translate messages from one protocol to another. All communications within the protocols are carried out using TCP/IP.

**Sensors:** The sensors will be devices that generate data for the cyber physical system being handled. They may use the TCP/IP protocol, to connect with a MQTT broker (extending the solutions to other messagin brokers is also possible). In most cases, the messages will be directed to the data processing component of the worker node.

### 4.1.3.3 Process View

The Level 3 Process View employs the UML sequence diagram notation to exemplify the design of the system use cases while additionally delving into the intricacies of the most captivating and intricate functionalities.

**UC1 - Login to the Web Interface**

Regarding user authentication in the web interface, the user will start by entering the necessary data for authentication. They will effectively enter the folder ID and the API key, where the former functions as an identification number and the latter acts as a password. After submitting the credentials, the web interface makes a request to the Edge4CPS API, which first checks if the POST ("HTTP Methods." Contrive.mobi, 2021) request fulfils all requirements and then, together with the local database, verifies the credentials. Finally, the request is returned to the interface, where it will either be authorized or not.

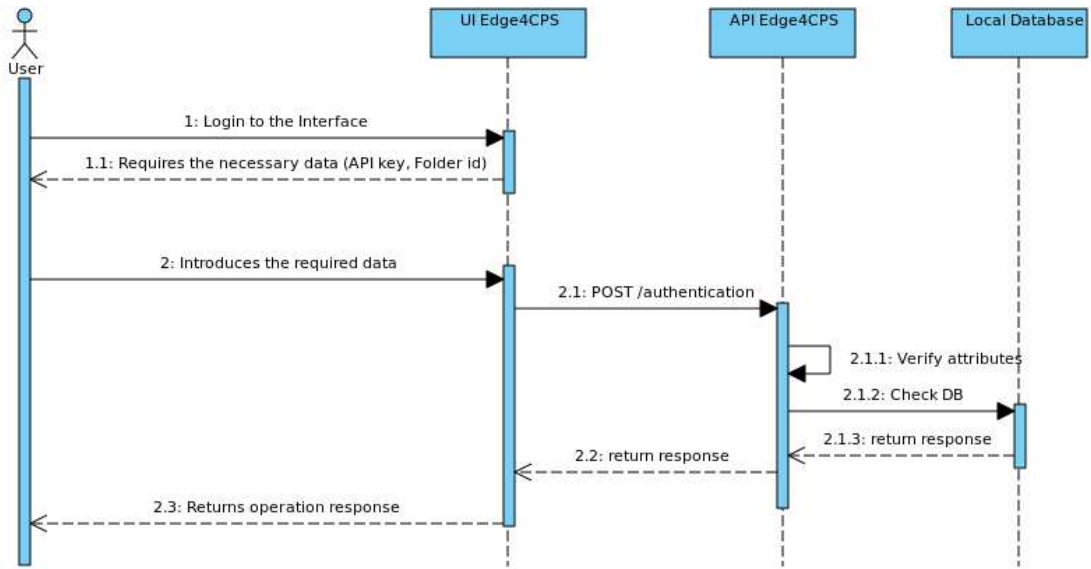

Figure 22 Sequence Diagram UC1: Login to the Web Interface

**UC2 - Deploy applications and services using the Web Interface**

For the deployment of applications and services by the user, they must be logged in. In Fig. 23, the login will not be represented as it has already been described in UC1. After user authentication, they navigate to the deployment window where they find a form to fill out. This form contains the container name, ports to be exposed, image name, CPU, and memory limits. After completing the form, it is submitted by the interface to the Edge4CPS system API through a POST request.

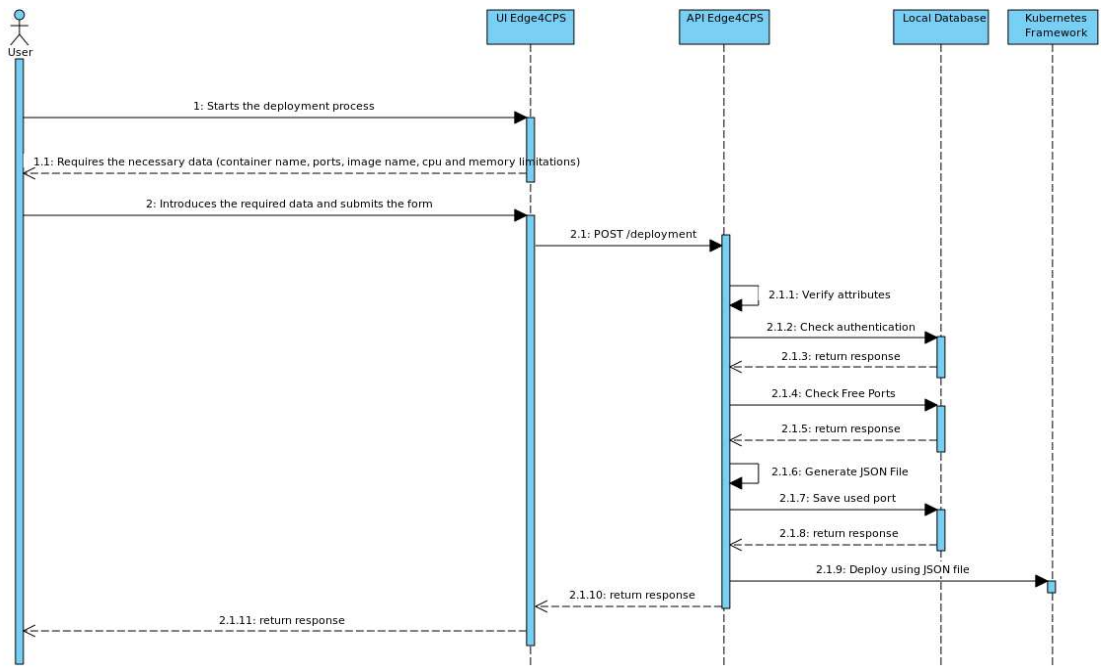

Figure 23 Sequence Diagram UC2: Deploy applications and services using the Web Interface

Upon receiving the request, the API verifies its integrity and checks the credentials against the local database. After this verification, it consults the database to determine which ports are available within the cluster for external exposure. Once it identifies the available ports, it generates the Kubernetes deployment file. Finally, the API initiates a deployment order using the deployment file within the Kubernetes framework.

**UC3 - Deploy applications and services using the Edge4CPS API**

The deployment of applications and services using the Edge4CPS API follows the same process and design as the previous UC2. Therefore, a sequence diagram will not be presented.

**UC4 - Check Services deployed, Exposed Ports**

Regarding the feature that allows a user to check which ports their services are exposed on; the user makes a GET ("HTTP Methods." Contrive.mobi, 2021) request to the Edge4CPS API. After receiving the request, the API verifies the integrity of the package and checks the credentials present in the request against the local database. Finally, it queries the database again to retrieve the corresponding exposed ports for the user's services, to then answer back to the request with the query.

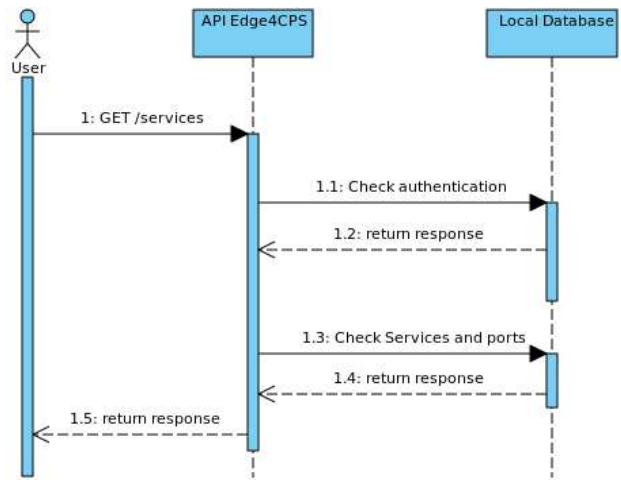

Figure 24 Sequence Diagram UC4: Check Services deployed, Exposed Ports

**UC5 - Register a New User**

To create a new user in the system, the system administrator makes a POST request to the Edge4CPS API, providing their own credentials as well as the credentials they wish to register for the user. When the API receives the request, it once again verifies if the request is well-formed and checks if the system administrator's credentials match those in the database. Finally, another request is made to the database to check if there is already a user with those credentials. If there isn't, the record of the new user is stored in the database.

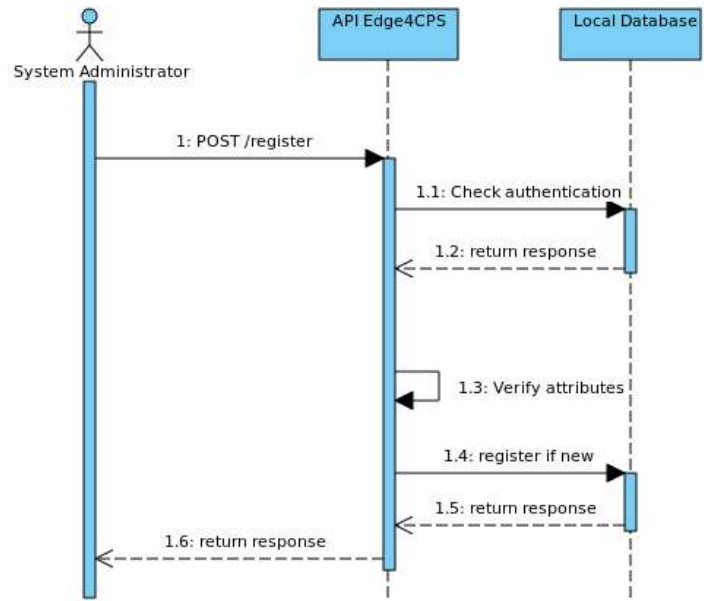

Figure 25 Sequence Diagram UC5: Register a New User

## UC6 - Make Operations Inside the Cluster

To allow the system administrator to perform operations within the cluster, they first need to connect to the master node via SSH ("Secure Shell (SSH).", 2021). After entering the username and password, the system verifies if the SSH credentials match the existing ones. If they do, the administrator gains access to a shell on the master node with access to the framework, enabling them to execute commands in the cluster.

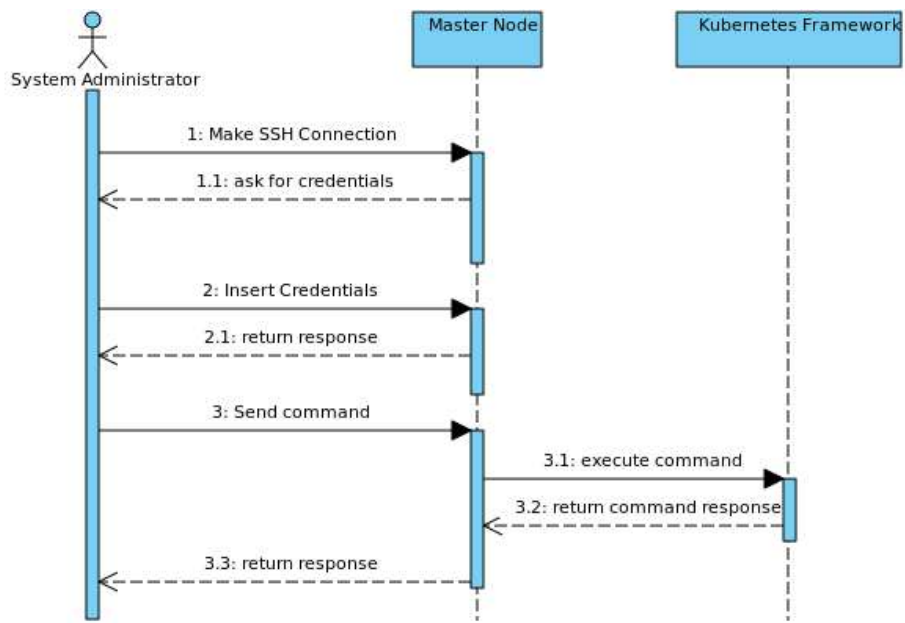

Figure 26  Sequence Diagram UC6: Make Operations Inside the Cluster

**UC7 - Check Cluster Components Status**

To allow the system administrator to check the cluster components, they first need to log in to the Kubernetes dashboard exposed by the cluster. After entering the credentials, the system will verify if they are correct. Upon successful authentication, the administrator will then be able to access and interact with essential Kubernetes components. The Kubernetes dashboard communicates with the cluster framework to retrieve information from it.

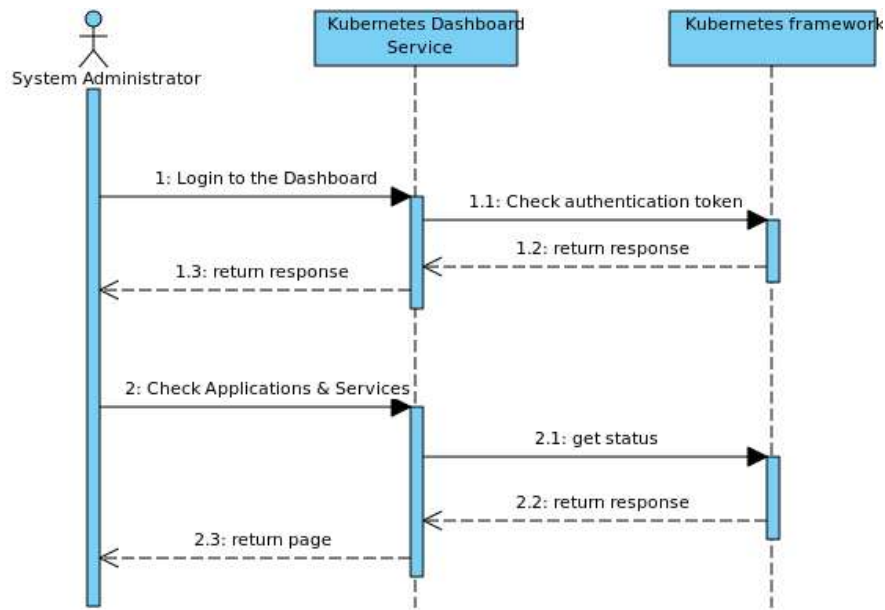

Figure 27 Sequence Diagram UC7: Check Cluster Components Status

**UC8 - Use the data pipeline implemented inside The System**

To utilize the data pipeline, the administrator will require a data producer that will generate data to the MQTT broker endpoint. Then, a subscriber within the data processing component will retrieve the data to be processed. After this step, the data is sent back to the MQTT broker, but this time the middleware will receive the data and translate it from MQTT to Kafka, producing the data for Kafka. On the other side, the classification component will receive the data from Kafka to implement classification algorithms on it. Finally, the data intended for visualization will be presented in an interface that provides the classified data and processed data for the administrator to view.

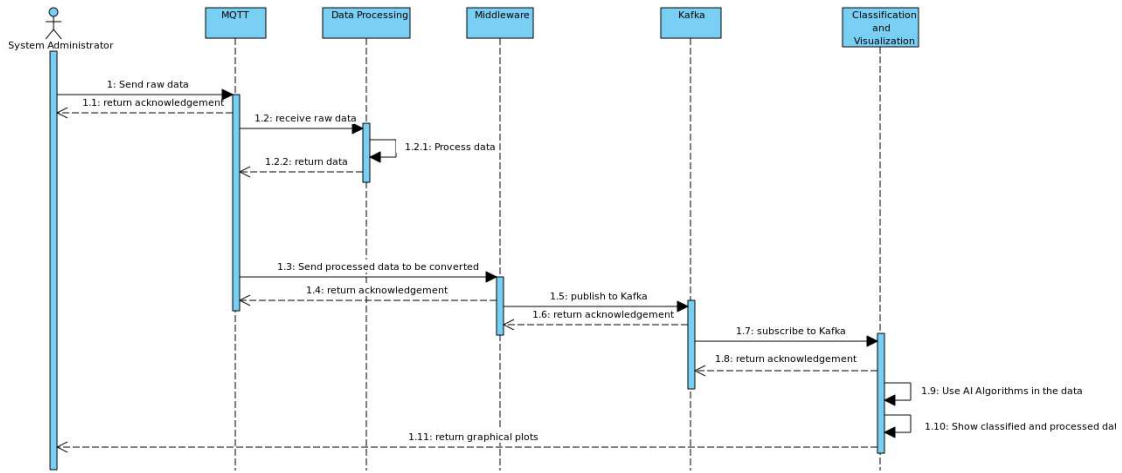

Figure 28 Sequence Diagram UC8: Use the data pipeline implemented inside The System

# 5  Implementation

In this chapter, we commence by delving into the technology choices made, providing insights into the decision-making process. Subsequently, an overview of the system setup is presented, offering a comprehensive understanding of its configuration. Lastly, the implementation of the solutions is elucidated using figures to demonstrate the system, and screenshots of its components.

## 5.1  Technologies

After studying various technologies for implementation in the Edge4CPS system, it was decided to implement the cluster using Kubernetes. This choice was made because it is the one that best aligns with our implementation. Thanks to the flexibility offered by the framework, Kubernetes provides a wide range of features and configurations that can be customized to meet specific application requirements. Another advantage is that our system requires easy scalability and high availability, both of which are offered by Kubernetes through its automatic scaling and load balancing capabilities (Burns & Tracey, 2018). Additionally, it automatically restarts failed containers, replaces unhealthy containers, and distributes workloads across available resources, thereby minimizing downtime and enhancing reliability.

For the containerized environment, it was decided to use Docker, as it will make it easier for users to utilize the Docker Hub to store Docker images.

In terms of the Edge4CPS API and the Data pipeline, Python was chosen and used due to its flexibility, extensive library support, and compatibility with various operating systems. Python is widely recognized for its versatility and is a popular choice for data processing tasks as well ("10 Reasons Why Python Is One of the Best Programming Languages.", 2022).

The Web Interface for Edge4CPS is developed using JavaScript, specifically React, due to its extensive capabilities. JavaScript is the most widely used programming language for web development, as it runs natively in web browsers, making it essential for client-side scripting. React, a powerful library known for its high performance and versatility, is an ideal choice for creating the Edge4CPS web interface (Islam Naim, 2017).

Finally, SQLite was chosen as the local database for several reasons. Firstly, SQLite is a lightweight and embedded database that does not require a separate server process to function. Additionally, SQLite works seamlessly with Python, making it a natural fit for the project's use of the Python programming language (Kreibich, J., 2010).

Visual Studio Code was utilized as the integrated development environment (IDE) for implementing solutions in Python and JavaScript programming languages. Its versatility and support for multiple programming languages made it an excellent choice for the development process ("Why Visual Studio Code?", 2023).

## 5.2  Edge4CPS System Setup

To set up the Edge4CPS system in a development environment, it is necessary to install a containerized environment on each machine that intends to join the cluster. In our system, the recommended choice is the Docker environment. After installing the containerized environment, it is important to install the Kubernetes framework, which should be installed after the containerized environment is set up. After installing the Kubernetes framework, the decision of what role the nodes will have, whether master nodes or worker nodes, will need to be made. On the master node, the Edge4CPS API and Web Interface will be installed, while the brokers and middleware will be installed on the worker nodes within the containerized environment. Finally, the computation pipeline will also run on the worker nodes.

## 5.3  System Overview

### 5.3.1  Cluster

The cluster is a crucial concept in Edge4CPS, implemented using Kubernetes. It is responsible for hosting, coordinating, managing, and scaling multiple computing nodes. As previously discussed, nodes within the cluster can serve two distinct roles: master nodes or worker nodes. The master node functions as the primary management machine and includes a Deployment Interface, an API, and a containerized environment, in Fig. 29 the master node components are shown.

On the other hand, the worker node typically only includes a containerized environment along with user applications and services. This setup increases the cluster's capacity to support additional applications and services.

As mentioned, the containerized environment exists on each node in the cluster, whether it be the master node or the worker node. Effectively, it is within this environment that chosen applications and essential support services run, including the environment monitoring dashboard, the users deployed services and applications, the messaging protocol brokers, and the middleware that enables the possibility of translating multiple communication protocols, as show on Fig. 29.

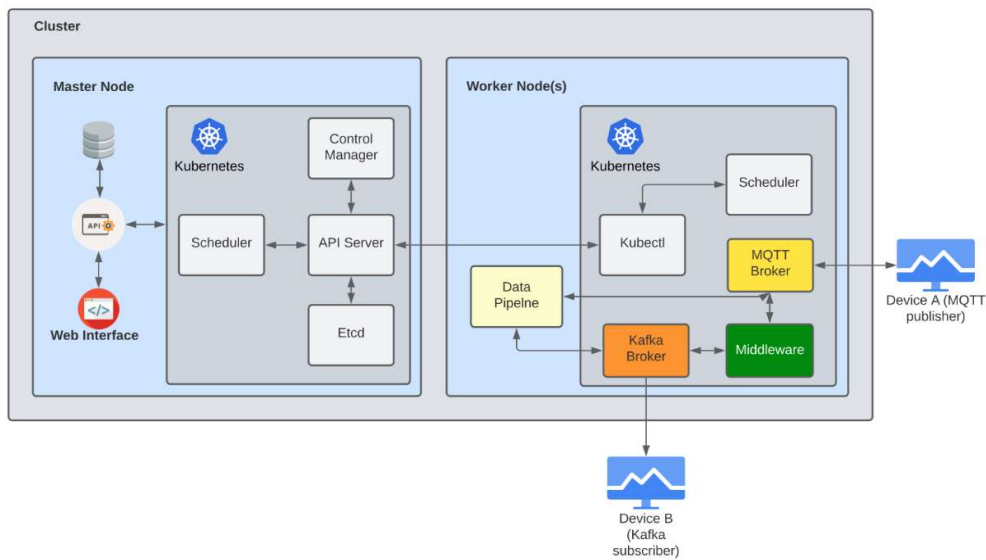

Figure 29 Global Cluster Components

### 5.3.2 Monitoring Components

The monitoring of the cluster is supported by the Kubernetes dashboard, as we see in UC 7.

This dashboard was only given access to the system administrator. Regular users can deploy applications and services using the web interface or the edge4cps API and verify the exposed services through the API. This ensures greater restrictions and security in access. Since it offers various monitoring capabilities, such as cluster and node monitoring, pod monitoring, resource usage monitoring, event tracking and logging, as well as the ability to monitor deployments and Replica Sets. Overall, the Kubernetes Dashboard offers a comprehensive set of monitoring tools for the cluster, making it easier to manage and troubleshoot all the applications and services, in Fig. 30 an overview of the Dashboard design ("Kubernetes Dashboards: A Comprehensive Guide.", 2022)

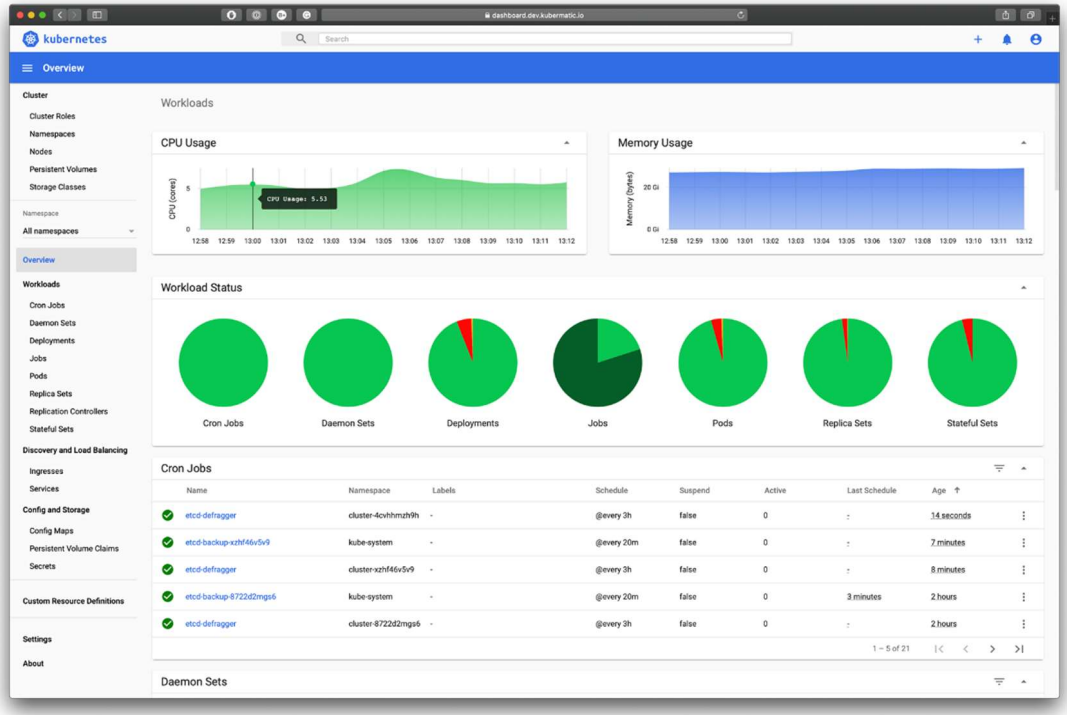

Figure 30 Kubernetes Dashboard

### 5.3.3   Edge4CPS UI

The Web Interface simplifies the user interaction with Edge4CPS, as it facilitates the deployment of applications and services.

The Deployment Interface was developed using React with the aim of abstracting the backend API and simplifying processes through a graphical representation. It includes an authentication process, Fig. 31, and allows users to fill out a form with information about the application to be deployed, Fig. 32.

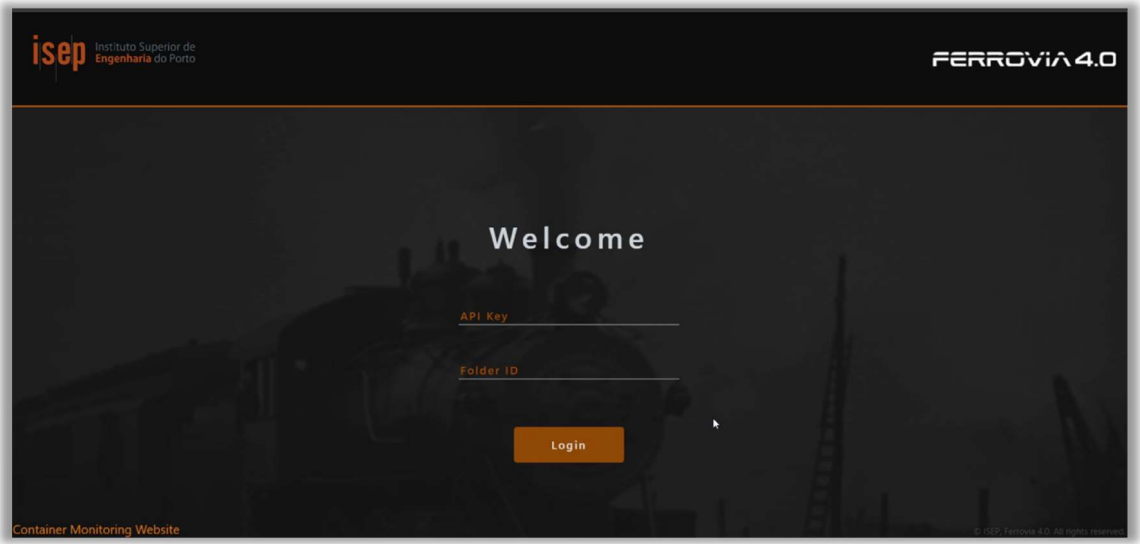

Figure 31 Edge4CPS Web Interface Authentication

The form from the Fig. 32, contains fields for basic configuration, such as, image name, container name and ports to be opened. Besides that, it also contains sliders, regarding to resource limits and needs, like CPU, and memory.

The API, which we will detail next, uses the form filled from the Deployment Interface to deploy the application to the cluster. Users can specify the image name and set resource limits for CPU, memory, and GPU cores. Furthermore, the Deployment Interface enables easy access to Kubernetes support applications, like to the dashboard for managing the containerized environment.

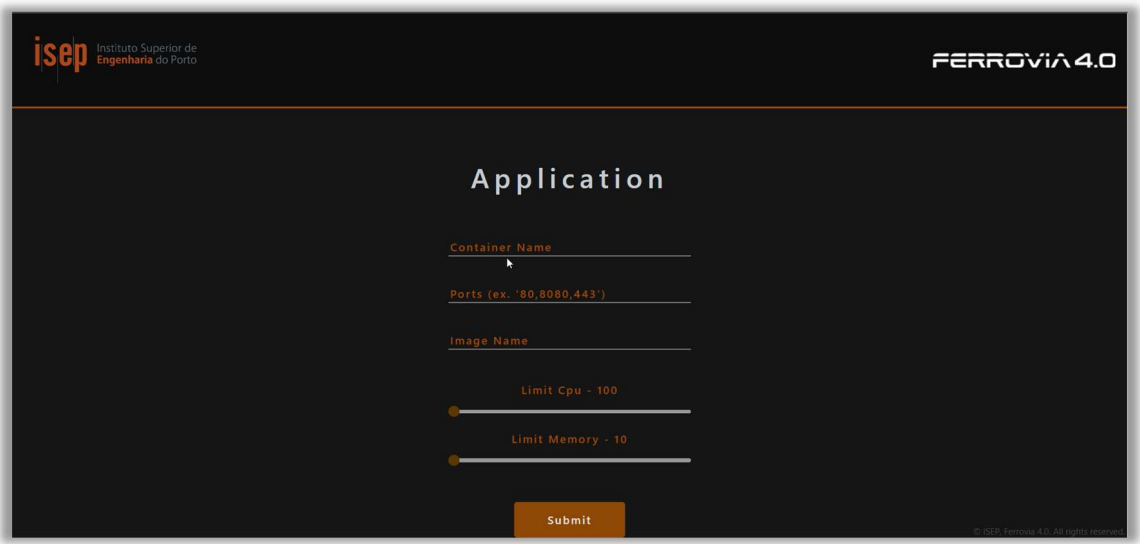

Figure 32 Edge4CPS Web Interface Deployment

### 5.3.4 Edge4CPS API

As mentioned previously, the API has been developed in Python and serves as the connection point between the user-oriented Deployment Interface and the cluster. Its responsibilities include creating a configuration file based on the specified deployment type and services by the user, and then using that file to deploy the service to the cluster nodes. The code snippet 1 shows the code used to generate the configuration file.

```python
def writeConfig_kubernetes(**kwargs):
    template = """apiVersion: apps/v1
kind: Deployment
metadata:
  name: {name}-deployment
  labels:
    app: {name}
spec:
  replicas: 1
  selector:
    matchLabels:
      app: {name}
  template:
    metadata:
      labels:
        app: {name}
    spec:
      restartPolicy: Always
      nodeSelector:
        node_type: {compatible}
      containers:
        - name: {name}
          image: {image}
          ports:
          - containerPort: {all_ports}
          resources:
              requests:
                  memory: "{memory}Mi"
                  cpu: "{cpu}m"
              limits:
                  memory: "{memory}Mi"
                  cpu: "{cpu}m"
"""

    with open(kwargs.get('file_name'), 'w') as yfile:
        yfile.write(template.format(**kwargs))
```

Code Snippet 1 Function to generate the Deployment File

To deploy a service on the cluster, the UI makes a REST ("What is a REST API?" Red Hat, 2020) request with parameters such as the access key and the name of the image, considering how it is named in the Docker Hub repository. The UI, as specified by the user, also includes the ports to be opened and the required resources and needs, for CPU, memory. It is also the API that, upon receiving the user's deployment request, automatically deploys the containerized applications based on the processor architecture that they were built upon. To do this the API checks what type of architecture the image supports, using its name, and then chooses the computing node that's most adequate for such application, which might be available for specific CPU architectures, such as AMD or ARM architectures.

### 5.3.5   Messaging protocol Brokers

In the Edge4CPS system, there will initially be two messaging protocol brokers, with the possibility of adding more later, if required. As messaging protocol brokers, we will have Kafka and MQTT, which will allow users to have applications running on the cluster that use these brokers as a means of communication, following the principles of the publish/subscribe paradigm. These brokers will also be used by two other components that will be discussed later: the middleware, which will be used for protocol translation, and the pipeline, which will use Kafka and MQTT to retrieve and publish messages.

### 5.3.6   Middleware

Edge4CPS middleware[1] aims to support multiple communication protocols in a scalable manner. It was developed as a configurable application that supports several messaging protocols, including Kafka, MQTT, and in the future others. This middleware is available in open-source code which can also be extended to support other messaging protocols.

The user can configure the middleware based on the publishers and subscribers that will be active, as publisher and subscriber in different middleware brokers. For example, the user could require an MQTT publisher to redirect messages to an AMQP and to a Kafka subscriber. The application will then create its own MQTT subscriber, and both an AMQP and Kafka publisher, connect the three of them internally based on additional configurations provided by the user (for example, which topic to publish to, or additional connection details), and the now running consumers will do the rest.

---

[1] I participated in the early stages of development and testing of the middleware, and later on, the project was handed over to Pedro Costa from the research group.

### 5.3.7   Data Pipeline

The data pipeline is a pre-built framework designed for easy and rapid deployment. Its components, such as data processing and classification, can be distributed across multiple CPU cores using threads and multiple nodes, as shown in Fig. 34. This allows for efficient utilization of computing resources and improved capabilities for data processing, cleaning, and feature extraction ("Parallel Programming: Definition, Benefits, and Best Practices.", 2023). In our implementation, we will combine the data cleaning and data acquisition components. In code snippet 2, the thread from the classification component is demonstrated.

```python
class ClassificationThread(QThread):
    def __init__(self,semaphore ,parent=None):
        super().__init__(parent)
        self.semaphore = semaphore
    def run(self):
        try:
            semaphore.acquire()
            classification(self.semaphore)
            semaphore.release()
        except:
            pass
```

Code Snippet 2 Thread Creation Classification Component

This pipeline was developed in Python, including the visualization part. In terms of the steps in the data pipeline, it starts with data production (on the sensors), which is later collected by the MQTT acquisition component, and then processed by the cleaning component. After that phase, the data is sent via MQTT to be translated into Kafka, as the classification component will subscribe to Kafka. When the data reaches the classification component, it will be passed through an AI algorithm for classification. Finally, the classified data, along with the processed data, will be presented to the system administrator in a visualization interface. Although the Edge4CPS system is independent, an implementation of it is within the scope of the Ferrovia 4.0 project, as mentioned before. The implemented data pipeline took into account the project requirements. Indeed, the data used in the pipeline relates to metrics measured by sensors distributed throughout the train. A demonstration of the data visualization component is shown in Figure 33, where the system was implemented inside a train at the Railway Summit 2023 ("Portugal Railway Summit.", 2023).

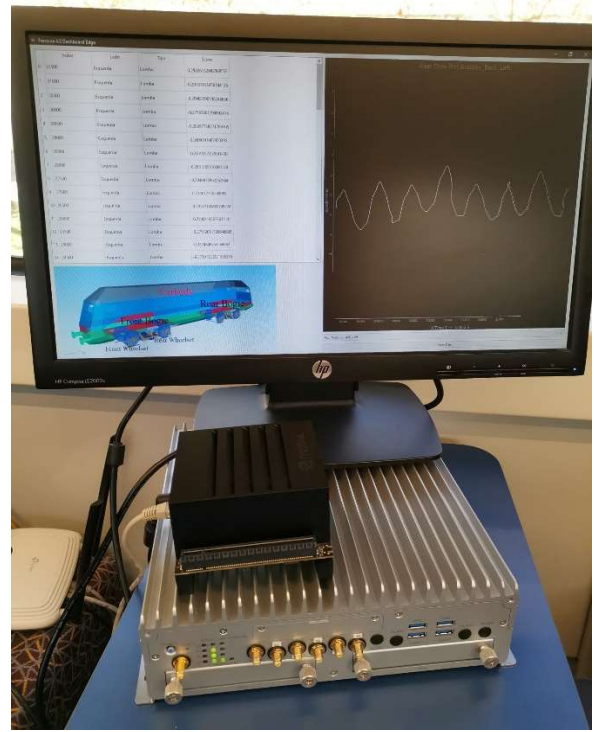

Figure 33 Data Visualization Component (Photographed at Railway Summit 2023)

With this pipeline, the system can be tested by the system administrator, who can use the visualization component for benchmarking and to check if anything went wrong in the process. The various components of the pipeline can be parallelized from the data processing component to the classification or the visualization component. In the pipeline implementation, both classification and visualization parts are effectively executed in multiple threads to ensure efficient and delay-free data collection, processing, and presentation. In Fig. 34, there is a figure that demonstrates all the components and how they interact with the system.

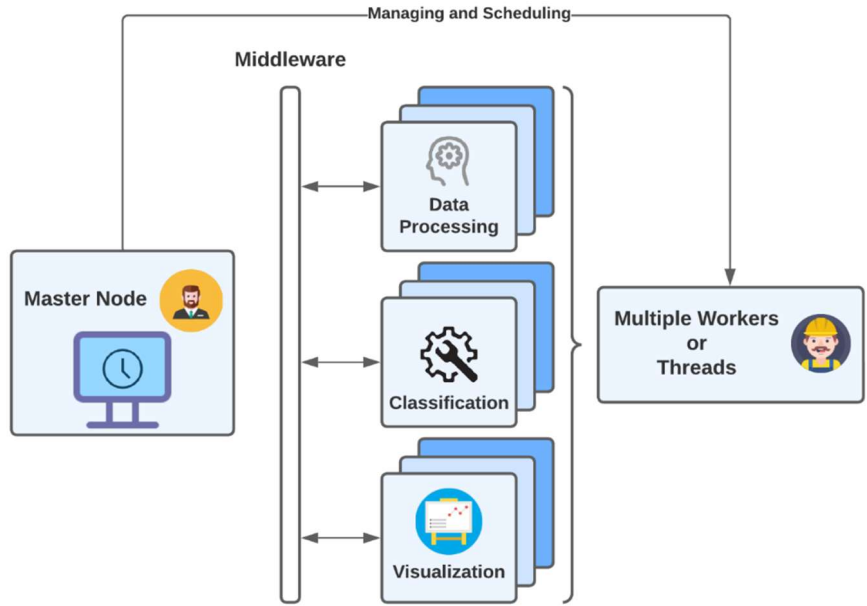

Figure 34 Edge4CPS Data Pipeline

### 5.3.8 Addressing

Cluster nodes (both master and worker nodes) can be connected to the same network. To allow services to be accessed from outside of the cluster, we can use Kubernetes' NodePort service or Ingress (Dinesh, 2018). Note that there is the need to configure firewall rules (of the operating system or network device on which the Kubernetes cluster is running) to permit incoming traffic to the cluster nodes, which is an effective way to control access and prevent unauthorized access. This will expose the network IP and forward traffic to the appropriate service port ("1-to-1 NAT Configuration Example.", 2023).

In Figure 35, users can see how to interact with the deployed services in the cluster. In this example, the cluster IP is 192.168.1.209. To communicate with other services and applications in the cluster, the user uses the IP, and the port exposed by the cluster. If an IoT device needs to send MQTT messages through the broker on the cluster, it needs to use the IP 192.168.1.209 and port 33012. For the Deployment Interface, the user would connect to IP 192.168.1.209 and port 8080, where the service is exposed.

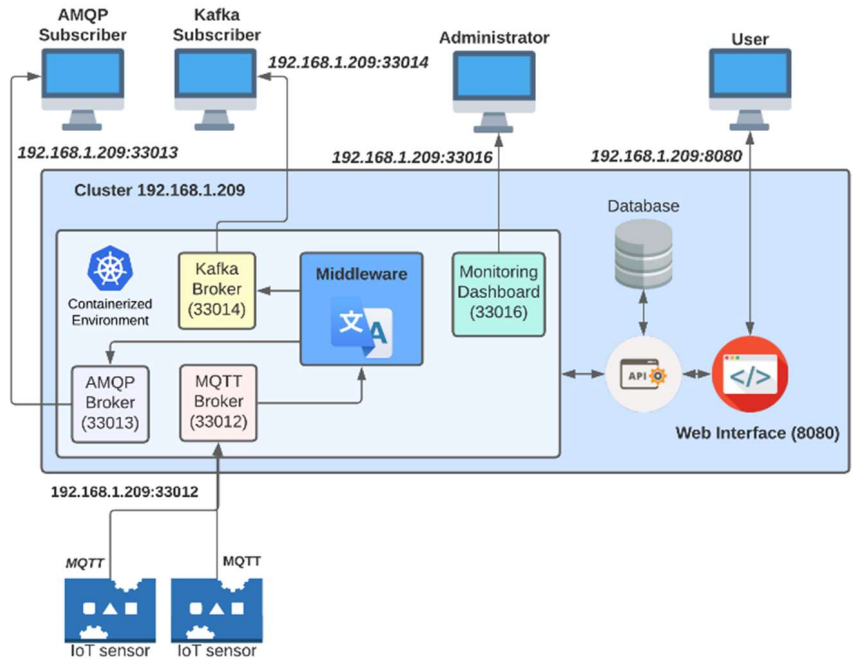

Figure 35 Edge4CPS Network example

# 6  Solution evaluation and tests

## 6.1  Tests

This section outlines the testing methods employed to validate the Edge4CPS system, ensuring the quality of the solution while meeting all requirements and constraints. The chapter covers three types of tests: unit tests, integration tests, and system tests. Lastly, an additional section will be included to discuss the utilization of CUDA cores in the data pipeline component to improve performance and present the benchmark tests conducted.

### 6.1.1  Unit

Unit testing focuses on testing individual units or components of a software system in isolation (Runeson, 2006). These units are typically small and self-contained, such as functions, methods, or classes. Unit tests verify that each unit behaves correctly according to its specifications.

Code snippet 3 demonstrates the unit-level tests performed on the API component. Some of these tests include verifying the method that accesses the Docker Hub and checks the user-provided image for compatibility with the supported CPU architectures. In addition to this test, it also executes tests related to generating ports to expose applications in the system, as well as generating the configuration file.

```python
class MainTaskTestCase(unittest.TestCase):

    def test_image_compatibility_right(self):
        result = main_task("nginx")
        self.assertEqual(result, "amd")

    def test_image_compatibility_bad_parameters(self):
        result = main_task("something-wrong")
        self.assertEqual(result, [])

    def test_port_text_beautify(self):
        result = port_text_beautify("8080")
        self.assertIn("8080|",result)

    def test_port_text_beautify_wrong(self):
        with self.assertRaises(AssertionError):
            result = port_text_beautify("")
            self.assertIn("8080|",result)

    def test_port_text_beautify_multiple(self):
        result = port_text_beautify('       - "10001:8080"        -
"10002:20022"')
        self.assertIn('|10001:8080"|10002:20022"|',result)

    def test_port_to_text_single(self):
        result = port_to_text([8080])
        self.assertEqual(len(result.split("-")),2)

    def test_port_to_text_multiple(self):
        result = port_to_text([8080,2020])
        self.assertEqual(len(result.split("-")),3)

    def test_port_to_text_empty(self):
        result = port_to_text([])
        self.assertEqual(result, "")
```

Code Snippet 3 Unit Tests API

## 6.1.2   Integration

Integration testing involves testing the interactions and interfaces between multiple units or components of a system (Delamaro, Maidonado, & Mathur, 2001). It validates the integration of these units and ensures that they work together as expected. Integration testing helps identify issues related to data communication, API integrations, or interoperability between different modules. The integration tests presented in Fig. 37 are related to the API component of Edge4CPS and will be used to verify the proper communication between the different components of the system through the API. Among the developed tests are the user registration request test, as well as the application and service deployment request test in the cluster, and the request to retrieve the exposed ports by the applications, which will return the ports that the user can use to communicate with the services. In Fig. 36, the code is used to test the request for exposed ports. It checks if the response is a JSON ("W3Schools - JSON Introduction.", n.d.) array and if the response code is 200, which indicates a successful HTTP response ("HTTP response status codes.", 2023).

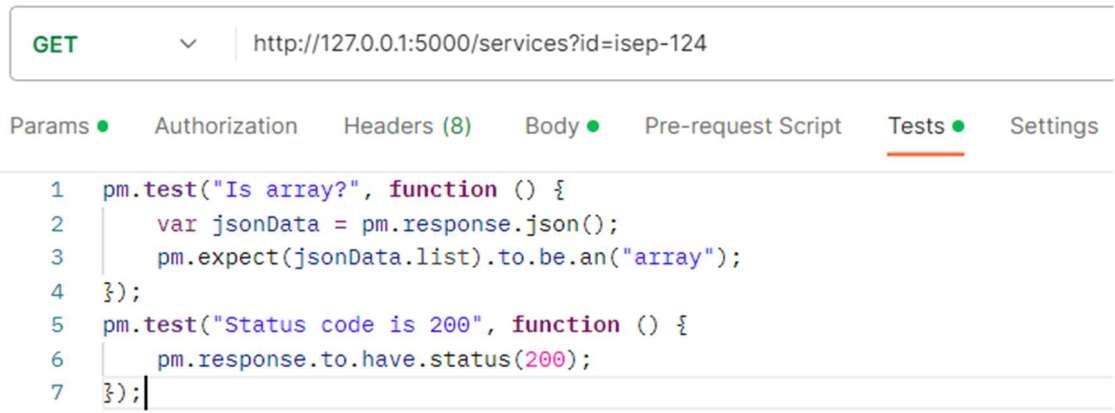

Figure 36 Integration Test Get Services Exposed Ports

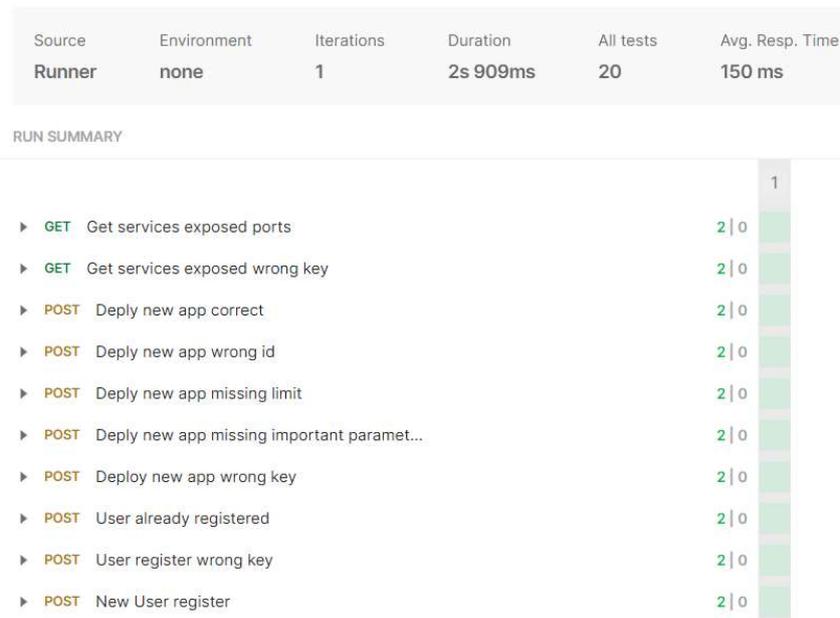

Figure 37 Integration Tests Postman

### 6.1.3 System Testing

System testing is a software testing methodology that verifies the complete flow of an application, starting from the system actor perspective and covering all the integrated components, subsystems, and dependencies ("System Testing.", n.d.). The goal of system testing is to ensure that the application behaves as expected and meets the requirements, simulating real-world usage scenarios.

### 6.1.3.1  Edge4CPS System

A test was conducted simulating the normal usage of the Edge4CPS system from the UI to the Kubernetes Framework. The test followed the following steps:

**Register User**

- **Job:** System administrator register user using the Edge4CPS API
- **Tested Components:** UI, API, Local database
- **Result:** The user was successfully registered, appearing inside the Database

**Login User UI**

- **Job:** User logs in to the UI
- **Tested Components:** UI, API, Local database
- **Result:** The user passed the authentication process

**Deploy de Aplicação na UI**

- **Job:** User deploys an application inside the cluster using the UI filling a form with the needed requirements
- **Tested Components:** UI, API, Local database, Kubernetes Framework
- **Result:** The user application is shown when requesting the applications deployed

**Verificação dos Componentes do Cluster**

- **Job:** System administrator uses the Kubernetes Dashboard
- **Tested Components:** UI, Kubernetes Dashboard, Kubernetes Framework
- **Result:** The system administrator was able to connect to the Kubernetes Dashboard and navigate through the dashboard.

### 6.1.3.2  Data Pipeline

The data used in the pipeline refers to metrics measured by sensors located on the train where the Edge is installed (Ferrovia 4.0,2023). Since it was not possible to run the system on a fully operational train with all the sensors, a mock scenario was created where the data used was previously measured on another train that was operating in a controlled environment.

The data generated in the controlled environment will be used in the data processing component of the data pipeline. This data will contain a data frame with multiple rows, each row representing the acceleration readings from 7 different sensors on the train. Upon receiving this data, the data processing component will publish 1 message through MQTT (corresponding to 1 row from the data frame) per millisecond, containing a combined measurement of the acceleration from the 7 sensors. Fig. 38 displays a view of the MQTT broker console after receiving 10 messages from the pipeline.

```
1686501634: Sending PINGRESP to auto-7E76BEBD-6A64-D6CF-6FC3-E45583F21E85
1686501694: Received PINGREQ from auto-7E76BEBD-6A64-D6CF-6FC3-E45583F21E85
1686501694: Sending PINGRESP to auto-7E76BEBD-6A64-D6CF-6FC3-E45583F21E85
1686501755: Received PINGREQ from auto-7E76BEBD-6A64-D6CF-6FC3-E45583F21E85
1686501755: Sending PINGRESP to auto-7E76BEBD-6A64-D6CF-6FC3-E45583F21E85
1686501815: Received PINGREQ from auto-7E76BEBD-6A64-D6CF-6FC3-E45583F21E85
1686501815: Sending PINGRESP to auto-7E76BEBD-6A64-D6CF-6FC3-E45583F21E85
1686501876: Received PINGREQ from auto-7E76BEBD-6A64-D6CF-6FC3-E45583F21E85
1686501876: Sending PINGRESP to auto-7E76BEBD-6A64-D6CF-6FC3-E45583F21E85
1686501936: Received PINGREQ from auto-7E76BEBD-6A64-D6CF-6FC3-E45583F21E85
1686501936: Sending PINGRESP to auto-7E76BEBD-6A64-D6CF-6FC3-E45583F21E85
1686501996: Received PINGREQ from auto-7E76BEBD-6A64-D6CF-6FC3-E45583F21E85
1686501996: Sending PINGRESP to auto-7E76BEBD-6A64-D6CF-6FC3-E45583F21E85
1686502057: Received PINGREQ from auto-7E76BEBD-6A64-D6CF-6FC3-E45583F21E85
1686502057: Sending PINGRESP to auto-7E76BEBD-6A64-D6CF-6FC3-E45583F21E85
1686502118: Received PINGREQ from auto-7E76BEBD-6A64-D6CF-6FC3-E45583F21E85
1686502118: Sending PINGRESP to auto-7E76BEBD-6A64-D6CF-6FC3-E45583F21E85
1686502178: Received PINGREQ from auto-7E76BEBD-6A64-D6CF-6FC3-E45583F21E85
1686502178: Sending PINGRESP to auto-7E76BEBD-6A64-D6CF-6FC3-E45583F21E85
```

Figure 38 MQTT Broker Console

In the classification component, as the data from the previous component is subscribed, whenever a threshold of 500 data points is reached, a new thread is created to perform the classification of those data points using AI algorithms. The purpose of the classification is to categorize the data as bumps or depressions and obtain the confidence level for each classification. After classifying the threshold of test data, they are sent to the visualization component via file sharing. In the visualization component, the unclassified test data will be presented in a plot, while the classified threshold data will be displayed in a separate table on the left, as show in Fig. 39.

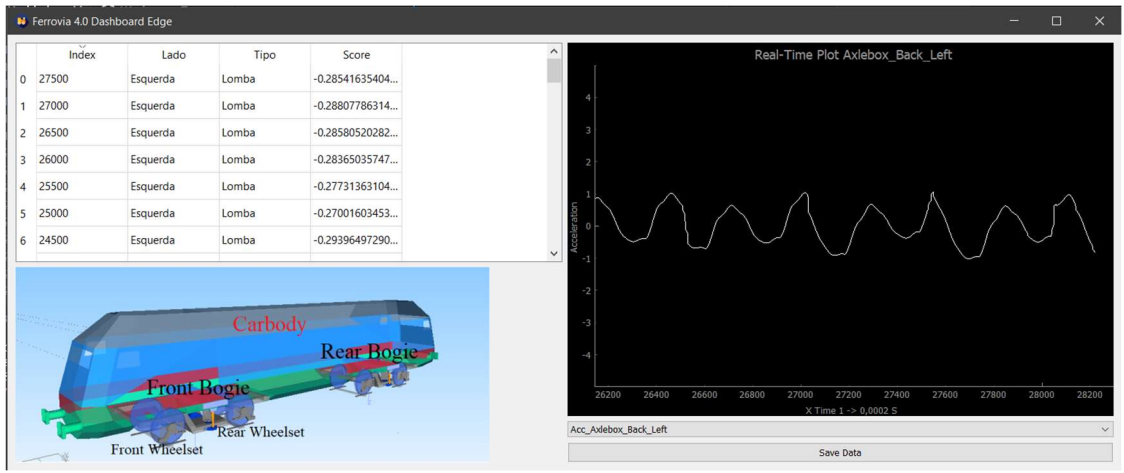

Figure 39 Data Pipeline Visualization Component (Ferrovia 4.0)

### 6.1.3.3  Middleware

The system tests conducted on the middleware followed the flow depicted in Fig. 40, where one device, in this case a PC, acted as a publisher while another PC device acted as a subscriber. The publisher used the MQTT messaging protocol to send a test message to the MQTT Broker, then the middleware subscribed to the message from the broker and published it to the Kafka broker. Finally, the subscriber test received the message using the Kafka messaging protocol, thus confirming the functionality of the system and the middleware itself.

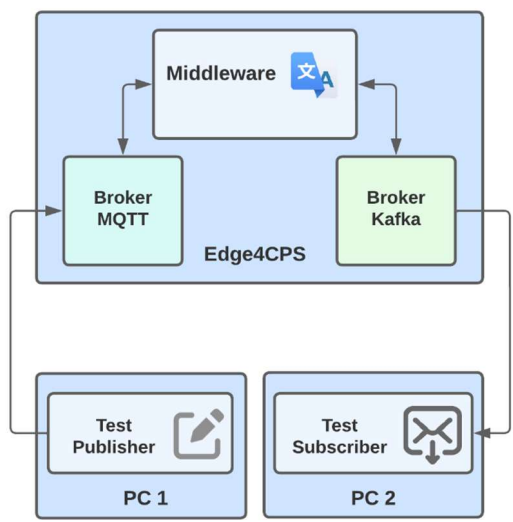

Figure 40 Middleware Test Flow

### 6.1.4 CUDA Cores

CUDA Cores are the processing units found in NVIDIA GPUs ("NVIDIA Tensor Cores.", n.d.). They are designed to efficiently execute parallel operations, enabling the acceleration of compute-intensive tasks. When programming in CUDA, it is possible to leverage CUDA Cores to accelerate compute-intensive operations. That's why the idea of implementing CUDA programming in applications arose, including the Python-based data pipeline of Edge4CPS.

Regarding the classification component of the data pipeline, part of it had already been implemented by consortium members. Consequently, the implementation of the algorithms and the most computationally demanding parts couldn't be done in CUDA. Despite this, a benchmark was conducted to assess the differences in processing speed.

The test application utilized the Numba library is a just-in-time (JIT) compilation tool for Python, which supports accelerating Python code on CPUs and can leverage the parallel processing capabilities of GPUs using CUDA. The benchmark involved testing three different compilation types: just-in-time compilation, CUDA compilation, and the normal Python interpreter (Oden, L., 2020).

The just-in-time compilation decorator instructs Numba to compile the function into optimized machine code just before execution. The CUDA decorator enables functions to run in parallel on the GPU, taking advantage of its massively parallel architecture and CUDA cores.

The first application was responsible for multiplying the values of two arrays with dimensions ranging from 100 to 10^8. Table 11 displays the results of the benchmark tests (execution time in seconds) for Application 1, separated by compilers Python, JIT and CUDA.

### Table 11 Benchmark Application

| Size | Python | JIT | CUDA |
|------|--------|-----|------|
| 10^2 | 0.0s | 0.1132s | 0.0644s |
| 10^3 | 0.0010s | 0.1137s | 0.0555s |
| 10^4 | 0.0020s | 0.1127s | 0.0602s |
| 10^5 | 0.0209s | 0.1177s | 0.0568s |
| 10^6 | 0.2503s | 0.1333s | 0.0677s |
| 10^7 | 2.4016s | 0.3412s | 0.1705s |
| 10^8 | 25.012s | 2.3477s | 0.5710s |

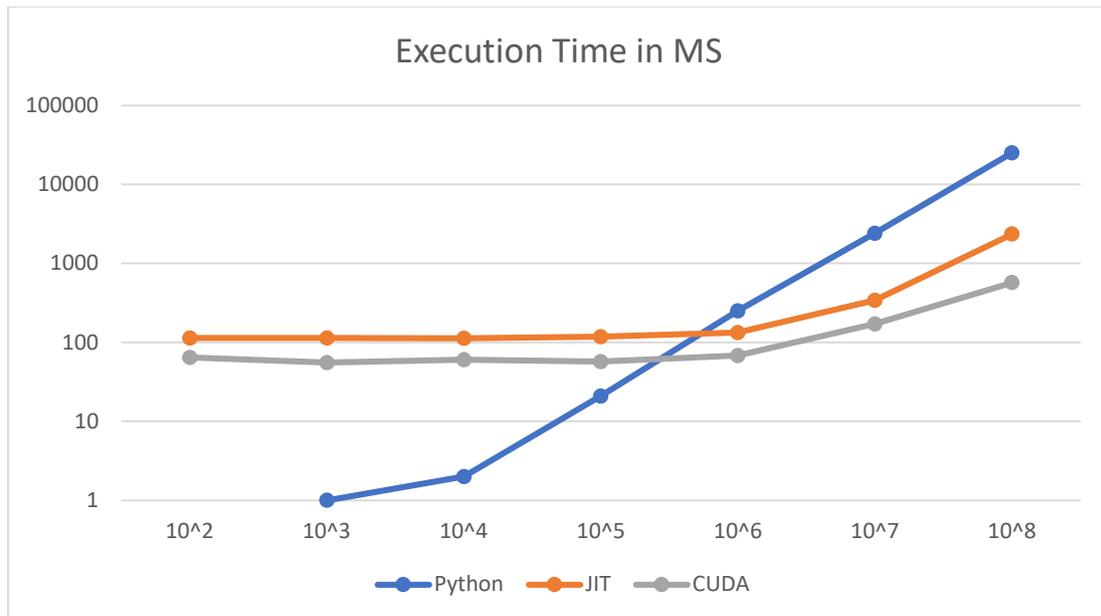

Figure 41 Plot Execution Time in Milliseconds

As observed in Table 11 and Fig. 41 as more demanding and costly operations are performed for the CPU, the Python interpreter struggles to keep up with the performance offered by the two other compilers.

# 7 Conclusions

This chapter provides an overview of the key aspects of this work, highlights the outcomes achieved through project development, elucidates the limitations and potential future enhancements of the project, and concludes with a final appreciation.

## 7.1 Summary

The goals of this project was to develop an open-source system that is easily scalable, manageable, and monitorable, allowing users to benefit from pre-built services within the system, such as a data processing pipeline, several communication middleware, and other services. Additionally, it aimed to provide users with the ability to easily deploy services and applications within the system.

First, a study of state-of-the-art solutions and technologies related to the project problem was conducted. This study allowed for a better understanding of the existing technologies in these field and which alternatives and hypothesis best fit the project. It also provided insights into different implementations within the same area and how they can differentiate themselves from Edge4CPS.

Furthermore, in Chapter 3, during the phase of requirements engineering and analysis, an examination was carried out to identify a comprehensive collection of both functional and non-functional system requirements. Also in that chapter, the domain model was presented, showcasing the system's structure and the relationships between its key elements. Additionally, the actors involved in the system were introduced and their respective roles and contributions were highlighted.

In the design chapter (Chapter 4), the focus was on the design principles, methodologies, and techniques utilized in the development of the system. It delved into the architectural design, outlining the overall structure and components of the Edge4CPS system.

In the implementation chapter (Chapter 5), the discussion primarily focused on the technologies utilized in each implemented component, their overall functionality, and the process based on the Ferrovia 4.0 use case implementation. The components covered include the master node, worker node, API, UI, Middleware, and Data Pipeline. Additionally, the chapter presented a general overview of the system setup process.

Regarding to the tests chapter (Chapter 6), it focused on the discussion of testing methods employed to validate the Edge4CPS system. Three types of tests were utilized: unit tests, integration tests, and end-to-end tests. The chapter also covered the utilization of CUDA cores in the data pipeline component, along with the presentation of a benchmark test on the same subject.

## 7.2 Accomplished goals

In relation to the initially proposed objectives, it is worth highlighting that all of them were successfully implemented, thoroughly tested, and thereby approved and considered completed. Table 12 summarizes the information by specifying a degree of accomplishment for each of the internship goals defined in the introduction section 1.1.1.

Table 12 Internship goals degree of accomplishment

| Goal | Degree of Accomplishment |
|---|---|
| Implement the system using Kubernetes with a focus on scalability, maintainability, and performance. | Complete |
| Ensure the system can support multiple applications and services with diverse requirements in a clustered environment. | Complete |
| Support the design, implementation, and test of a middleware, within the cluster's containerized environment, that is capable of translating messages between multiple messaging protocols, like Kafka and MQTT. | Complete |
| Design, develop, and test an API that can handle requests to the Kubernetes Framework within the master node, making it easier to deploy applications and services. | Complete |
| Develop a user-friendly web interface that implements the existing API, simplifying its complexity and offering users a graphical interface to manage the deployment of applications and services. | Complete |
| Establish a data pipeline within the cluster that enables efficient processing, handling, classification, and visualization of data. | Complete |

## 7.3 Limitations and future development

Although the project achieved all the initially planned goals, it is possible to exemplify some limitations that occurred, as well as provide some examples of future features, modifications, or extras.

- Although all system components have been tested and the pipeline has been tested with mock data from the Ferrovia 4.0 project, an even more realistic simulation where the system is implemented in a real-life scenario would be ideal.

- In the testing chapter, it is mentioned that the algorithm used in the classification component is not implemented in CUDA, which could potentially serve as a starting point in the future to further enhance the performance of the pipeline.

- Regarding the middleware, for greater flexibility in the future, support for additional messaging protocol brokers could be implemented.

- For future plans, the Edge4CPS UI could support the deployment of services and applications through templates, offering users even greater flexibility and simplicity in utilizing the system. This would enable users to quickly deploy applications by abstracting the process further. While efforts have already begun to implement this functionality in the API, it has not been integrated into the UI yet due to the need for further testing and validation.

- Another issue with the UI is that, for deploying applications and services, it could benefit from supporting a wider range of options and dynamic configurations by allowing JSON in the deployment configuration. This would provide users with the flexibility to either use the existing form to fill in deployment options or utilize JSON for more extensive configuration choices.

- For future plans, it will be interesting to support custom QoS in the deployment of applications and services within the system.

## 7.4  Additional Work Done

During the project development, a contribution was made to the INDIN 2023 conference, where the paper titled "A Scalable Clustered Architecture for Cyber-Physical Systems" was published. It is worth mentioning that this report received contributions from the same paper. Additionally, the project was also showcased at the Railway Summit 2023 event. In addition to these contributions, I also participated in the initial prototype development and testing of the middleware. Furthermore, throughout the project, I supported the Critical Computing Systems Open Day and attended various lectures related to the project topic. Finally, being part of the SoftCPS research group, I actively engaged in dialogues and contributed to finding solutions for the OPEVA, ARROWHEAD, and SailShape projects.

## 7.5  Final appreciation

Considering the implementation of the Edge4CPS system, including the UI, API, master node, worker nodes, middleware, and data pipeline, and after a brief analysis and testing, it can be concluded that the project has been successful, as everything proceeded as planned. Furthermore, the successful implementation at the Railway Summit 2023 further reinforces the project's positive outcome.

Despite meeting the initial expectations, it is important to consider the points presented in subchapter 7.4 (Limitations and Future Development), where improvements and changes to the system were proposed to make it more dynamic, flexible, and optimized.

On a personal level, working on a research project in collaboration with a consortium of various companies and universities was a fantastic and enriching opportunity. It provided me with the chance to learn about the DevOps aspect, which I hadn't been previously exposed to. Additionally, it gave me the opportunity to present this project at the Railway Summit 2023, and it also opened doors for me to present it at the INDIN 2023 conference. This experience contributed greatly to both my technical and social skills development.